\documentclass[1p]{elsarticle}

\usepackage[colorinlistoftodos]{todonotes}

\RequirePackage{rotating} 
\usepackage{subcaption}
\usepackage{graphicx}

\usepackage{booktabs}
\usepackage{multicol}
\usepackage{multirow}

\usepackage{amsmath}
\usepackage{amsfonts}
\usepackage{amssymb}

\usepackage{lineno,hyperref}
\usepackage{color}

\usepackage{csquotes}
\usepackage{comment}

\usepackage{natbib}
 \newcommand\citeapos[1]{\citeauthor{#1}'s (\citeyear{#1})}
 
\journal{arXiv}

 \biboptions{authoryear}

\begin{document}

\begin{frontmatter}


\title{Option market (in)efficiency and implied volatility dynamics after return jumps}
\author[mymainaddress]{Juho Kanniainen\corref{mycorrespondingauthor}}
\cortext[mycorrespondingauthor]{Corresponding author}
\ead{juho.kanniainen@tut.fi}

\author[mymainaddress]{Martin Magris}
\ead{martin.magris@tut.fi}

\address[mymainaddress]{Laboratory of Industrial and Information Management, Tampere University of Technology, P.O. Box 541, FI-33101 Tampere, Finland.}

\begin{abstract}
In informationally efficient financial markets, option prices and this implied volatility should immediately be adjusted to new information that arrives along with a jump in underlying's return, whereas gradual changes in implied volatility would indicate market inefficiency. Using minute-by-minute data on S\&P 500 index options, we provide evidence regarding delayed and gradual movements in implied volatility after the arrival of return jumps. These movements are directed and persistent, especially in the case of negative return jumps. Our results are significant when the implied volatilities are extracted from at-the-money options and out-of-the-money puts, while the implied volatility obtained from out-of-the-money calls converges to its new level immediately rather than gradually. Thus, our analysis reveals that the implied volatility smile is adjusted to jumps in underlying's return asymmetrically. Finally, it would be possible to have statistical arbitrage in zero-transaction-cost option markets, but under actual option price spreads, our results do not imply abnormal option returns.
\end{abstract}


\end{frontmatter}



\section{Introduction}

The arrival of a jump in an underlying price is important news for option traders. If financial markets are informationally efficient, option-implied volatility (IV) should immediately and non-gradually be adjusted  to new information that arrives along with a return jump for two reasons: (i) the IV stems from the option quotes that reflect option traders' conditional expectations of option payoffs and (ii) the IV accurately approximates a conditional (risk-neutral) expectation of cumulative future realized volatility \citep[see, e.g.,][]{carr2006tale}. According to the extant literature on return and volatility jump models, there are co-jumps in price and volatility, often in the opposite direction \cite[see, for example][]{eraker2004stock,jacod2010price,todorov2011volatility,bandi2016price}.\footnote{The negative relationship between price changes and volatility is often related to the leverage effect or the volatility feedback effect \cite[see, for example,][and references therein]{bollerslev2006leverage,kanniainen2013stock}. The difference lies in causality: the leverage hypothesis contends that return shocks lead to changes in volatility, whereas volatility feedback effect theory predicts that changes in volatility leads to return shocks. Moreover, there is also evidence that volatility affects stock returns through its impact on liquidity provision \citep{chung2018market}. In this paper, we consider how return jumps affect {\em implied volatility} through option pricing.}
However, results on simultaneous jumps in price and volatility do not rule additional delayed, gradual movements in the IV out, which would contradict the efficient market hypothesis. 

In this paper, we examine whether S\&P 500 index option markets react to a {\em single} jump in the underlying price causing {\em multiple} directed movements in IV so that the IV level is driven in a certain direction gradually within the next trading hour after the return jump arrival, which would provide evidence contrary to the efficient market hypothesis. Option markets are particularly interesting when studying market efficiency because, compared to stocks, option contracts and their payoff structures are precisely defined in terms of underlying variables. Thus, options can be priced at extremely high frequencies using precise mathematical models with high computational power. With advanced mathematical models and automated trading algorithms, such option contracts should immediately reflect the arrival of new information in modern electronic markets. 

Our analysis is performed as follows. Corresponding to event studies that analyze cumulative stock returns around news arrivals, we analyze minute-by-minute cumulative changes in implied volatilities within a one-hour post-window after the arrival of a return jump. We first perform a regression analysis where the IV movements that appear within the return jump interval are included (i.e., IV movements that are simultaneous with the return jumps), and we then run another regression by excluding them. Hence, we can analyze the total impact of a return jump on the IV, as well as the post-reactions in the IV to a recent return jump. Moreover, the heteroscedasticity of volatility is controlled in the regressions. The regressions are run not only for at-the-money options but also for the other moneyness groups using principal components extracted from the IV smile data for three, six, and nine months maturities. In our analysis, we extract at-the-money IV (ATM-IV)  smile curves with various maturities for every minute over a five-year period from the beginning of 2006 to the end of 2010. Regarding jumps in underlying prices, several jump detection methods are available, including \cite{barndorff2004power,lee2008jumps,ait2009testing}. This research relies on \citeapos{lee2008jumps}'s well-established non-parametric test. Because the jumps are non-uniformly spread across the trading hours, this research focuses on the first trading hour, where the vast majority of jumps occur. 

By using this simple testing scheme, we provide strong evidence that there are delayed, gradual movements in the IV within the next few minutes after the jump arrival. The post-jump IV movements drive the IV further down (up) after positive (negative) return jumps. In particular, the results are highly significant for IVs extracted from at-the-money options and out-of-the-money-puts (in-the-money-call), while the out-of-the-money call (in-the-money-put) IV converges to its new level immediately within the return jump window. This indicates that the implied volatility smile is adjusted to jumps in underlying's return asymmetrically. Second, regarding at-the-money options and out-of-the-money-puts (for which we obtain significant results), we find that the positive post-changes in the IV on negative return jumps continue during the following trading hour; however, negative post-changes on positive return jumps vanish within 60 minutes. We also visualize average IV dynamics after return jumps to demonstrate these findings. Although the opposite large movements in underlying prices and volatility are already reported in the extant literature \cite[see, for example,][]{eraker2004stock,bandi2016price}, our finding that negative (positive) movements in IV, which can be either large or small, tend to gradually increase (decrease) the IV level after a single positive (negative) jump in underlying price is completely new.\footnote{It is irrelevant if these IV movements are very large, representing 'jumps' in volatility, or are of smaller magnitudes; the point is that the IV changes to a new level gradually rather than immediately.} Remarkably, our results indicating that index option markets are not informationally efficient hold for the S\&P 500 index, which is based on liquid stocks, and its index options are among the most important stock derivatives. 

In terms of economic significance, we find that, compared to the situation where there have been no return jumps, the at-the-money implied volatility increases by an additional 0.218\% {\em after} the arrival of a negative jump over the next 60 minutes. In terms of call option dollar prices, this can correspond to a 1\% return or more, which would be a good return for a 60-minute period. While it would be possible to have statistical arbitrage in zero-transaction-cost option markets by trading at mid-prices, our results do not imply abnormal option returns under the actual bid-ask spreads in option markets. Overall, we find that (i) option markets are not informationally efficient because they adjust the implied volatilities gradually, but  (ii) no abnormal returns are observed when the transaction costs are considered. 

This paper considers two potential explanations for our finding: (i) informationally inefficient option markets and (ii) informationally inefficient stock markets for the underlying securities. The first explanation is considered more plausible, although the two are not mutually exclusive.
\begin{itemize}
\item [] {\em Informational inefficiency in option markets.} The first explanation is that option markets are inefficient if they do not fully incorporate new information regarding expectations of option payoffs at once, but gradually. This  is related to \citep{stephan1990intraday}, which states that ``in perfectly functioning capital markets, [...] new information disseminating into the marketplace should be reflected in the prices and the trading activity of both [option and underlying] securities simultaneously.'' Therefore, if the option markets gradually adjust the IV in a certain direction in response to a single jump in the underlying price, the option markets can be considered to be informationally inefficient.
\item [] {\em Informational inefficiency in stock markets.} The second explanation relates to the efficiency of underlying stock markets and the properties of spot volatility dynamics rather than the efficiency of option markets. \cite{jones1998macroeconomic} argues that the strong volatility persistence of announcement shocks indicates that some feature of the trading or information-gathering process itself causes volatility to be autocorrelated. Given that underlying price jumps represent information arrivals for traders in stock markets \citep{lee2011jumps,kanniainen2017arrival} and that at-the-money IV serves as a good proxy for spot volatility \citep{ai2007maximum}, the interpretations of \cite{ederington1993markets} are closely connected to our findings: the IV---and thus, the actual (unobservable) spot volatility---is not driven to a certain level at once, but gradually, which can indicate stock market inefficiency. 
\end{itemize}

While it is difficult to isolate the source of observed IV dynamics after underlying return jumps and answer whether the observed gradual post-movements in the IV are caused by option or stock market inefficiency or both, we believe that the main source of our findings is the inefficiency of option markets (the first explanation). The reason is that, according to the second explanation, option traders should be able to observe immediate changes in actual spot volatility on a tick-by-tick basis and incorporate this information into option prices immediately. However, spot volatility is an unobservable variable, and its robust estimation \cite[see, for example,][]{andersen2003modeling,mcaleer2008realized} has been a non-trivial task in the high-frequency econometric literature, especially because, under microstructure noise and stochastic volatility, ``market microstructure noise totally swamps the variance of the price signal at the level of the realized variance'' \citep{ait2005often} \cite[see also][]{zhang2005tale}. Even if daily realized volatility could be estimated reliably using intraday data, it is needless to say that it is extremely challenging, if not impossible, for practitioners to immediately obtain reliable tick-by-tick estimates of realized volatility using tick-by-tick time-series data for stock prices in a real-life online trading environment.\footnote{It should be noted that it would not be a good idea for option traders to extract tick-by-tick realized volatility estimates from option implied volatility (Black-Scholes IV or VIX) to price and trade options accordingly. This is because, by trusting the IV or VIX, informed option traders would consider the market prices of options to be truly correct, in which case, there would be no informational reasons to trade options!}

Our paper is a part of the literature on behavioral finance that questions market efficiency. So far, the literature has mainly focused on stock market efficiency \citep{shiller1981stock,de1985does,campbell1987cointegration}, but some research on option market efficiency exists. Some papers, such as \citep{harvey1992market,cavallo2000empirical}, argue against market inefficiency, which is partially in line with our finding that abnormal returns do not systematically exist in option markets. At the same time, some of the existing results provide evidence about possible inefficiency in option markets \cite[see][and references therein]{poon2000trading,ackert2001efficiency,deville2007liquidity}. Moreover, \cite{bernales2016investors} find that option investors tend to herd on dates of macroeconomic announcements, among other kind of periods of market stress. Because return jumps are related to macroeconomic announcements \citep{lee2011jumps}, our results are closely related to \citep{bernales2016investors}. Moreover, \cite{stephan1990intraday} find that the stock market leads the option market both in terms of price changes and trading activity, which is related to our results about informationally inefficient option markets (despite tha lack of systematic abnormal returns). In contrast to \citep{stephan1990intraday}, we examine the consequences of return jumps and not just any price changes; our analysis is model-free; and we consider index options rather than equity options. The reason for focusing on return jumps instead of price changes with any magnitude is that such large movements should lead to immediate and observable price discovery as they represent important news in option markets. Notably, our data is from 2006 to 2010 on the S\&P 500, during which index options can have been traded by algorithms technically enabling efficient price discovery. 

Secondly, this paper is part of the literature that examines IV dynamics in general. Because we want to run the analysis not only for ATM-options but also across the moneyness groups, we have to characterize IV smiles reliably. In the existing literature, methods for characterizing IV smiles or surfaces rely on refinements of principal component analysis (PCA) \citep{skiadopoulos2000,cont2002stochastic,fengler2006}. To examine at-the-money IV, as well as the entire IV smile, we use \citet{skiadopoulos2000}'s method to extract the principal components of IV smile dynamics. In line with \citep{cont2002}, \citep{fengler2003}, and \citep{balland2002}, in our analysis, the first three components capture most of the variability in the surface and are of clear interpretation. The extant literature on IV smile dynamics has typically used daily data, but not around underlying price jumps with intraday data nor associating the results to market efficiency, which we do in this paper. 

The structure of the paper is as follows. Section \ref{SEC:ImpVol} reviews the concept of implied volatility from the viewpoints of option prices and the risk-neutral expectation of future cumulative volatility. Section \ref{SEC:Data} introduces the datasets used in this paper (i.e., both option data and return jump locations that are detected from tick-by-tick time-series data). Section \ref{SEC:Meth} discusses the methods used in the paper, and Section \ref{SEC:Results} presents the results. Finally, Section \ref{SEC:Conclusions} concludes and discusses.

\section{Implied volatility}\label{SEC:ImpVol}

Since the introduction of the Black-Scholes theory, study and understanding of the IV has been a major area of effort for financial econometrics \citep{gatheral2011volatility}. Although the Black-Scholes model assumes that return volatility is constant, it empirically varies with respect to strike price $K$ and time to maturity $\tau$. Among early studies, \cite{rubinstein1994implied} finds smile features in the Black-Scholes IVs for S\&P 500 index options, while \cite{xu1994magnitude} find the same features in the Philadelphia Exchange foreign currency option market, and \cite{heynen1994empirical} in the European Options Exchange.  

The smile and term structure features are merely cross-sections of the so-called implied volatility \textit{surface}, IVSF, that jointly describes the relationship between the IVs with different strikes and maturities for a given period. The IVSF can be defined as 
\begin{equation}\nonumber
\sigma_{t}^{BS}: (K/S_t,\tau_t) \rightarrow \sigma_{t}^{BS}(K/S_t,\tau_t),
\end{equation}
by mapping a point $(K/S_t,\tau_t)$ to a point on the surface $\sigma_{t}^{BS}(K/S_t,\tau_t)$ such that
\[
C^{BS}(S_t, K, \tau_t, \sigma^{BS}_t(K/S_t, \tau_t)) = C_t^*(K/S_t, \tau_t),
\]
where $C^*_t(K/S_t, \tau_t)$ denotes the market price of an option with strike $K$, underlying price $S_t$, and maturity $\tau_t$ \cite[see, for example,][]{cont2002}. This expression reveals the three dimensions in which IV varies: time $t$, strike price scaled by the spot price at time $t$, $K/S_t$, and time-to-maturity $\tau_t = T-t$, where $T$ is the maturity time. Variations in the dimensions of moneyness and time to maturity are referred to as IV statics, whereas variations in the time dimension $t$ are referred to as IV dynamics. 

According to \cite{carr2006tale}, ATM-IV is an accurate approximation of the conditional risk-neutral expectation of the return volatility over the time horizon from the current time to the maturity time of an option(s):
\begin{equation}\nonumber
\mathbb{E}_t^Q \left(\sigma_{t,T}\right) = \sigma^{\text{ATMBS}}_{t,T} + O((T-t)^{\frac{3}{2}}),
\end{equation}
where 
\[
\sigma_{t,T} \equiv \sqrt{\frac{1}{T-t}\int_t^T \sigma^2_s ds}
\]
is the annualized realized volatility over $[t, T]$, while $\sigma_s^2$ is the (unobservable) spot squared volatility (variance) at time $s$, and $\sigma^{\text{ATMBS}}_{t,T} \equiv \sigma_{t}^{BS}(1, T-t)$ is ATM-IV with maturity time $T-t$. Moreover, $\mathbb{E}^Q(\cdot)$ refers to the risk-neutral expectation. Consequently, ATM-IV has a clear connection to spot volatility dynamics. Importantly, this relationship holds only if option markets are efficient, in which case, IV should react on changes in volatility expectations immediately and not gradually. Therefore, if the arrival of an underlying price jump affects volatility expectations, in perfectly efficient capital markets, the new information should be reflected in both the underlying price and option-implied volatilities simultaneously.

\section{Data}\label{SEC:Data}
\subsection{Intra-day option data}
We analyze the dynamics of ATM-IV and volatility smiles implied from the market prices of calls and puts on the S\&P 500 index, SPX. IV and smiles are analyzed with maturities of three, six, and nine months. Our choice of dealing with these maturities is motivated by the fact that longer maturity ranges correspond to lower liquidity scenarios. The dataset studied contains the spot prices and cross-sections of put and call prices for SPX in one-minute intervals and spans a five-year period from the beginning of 2006 to the end of 2010 (1,259 trading days in total). The option data is provided by CBOE Livevol, and we use daily interest rates provided by Optionmetrics. 

\begin{figure}[!h]
\centering
\includegraphics[scale=0.45]{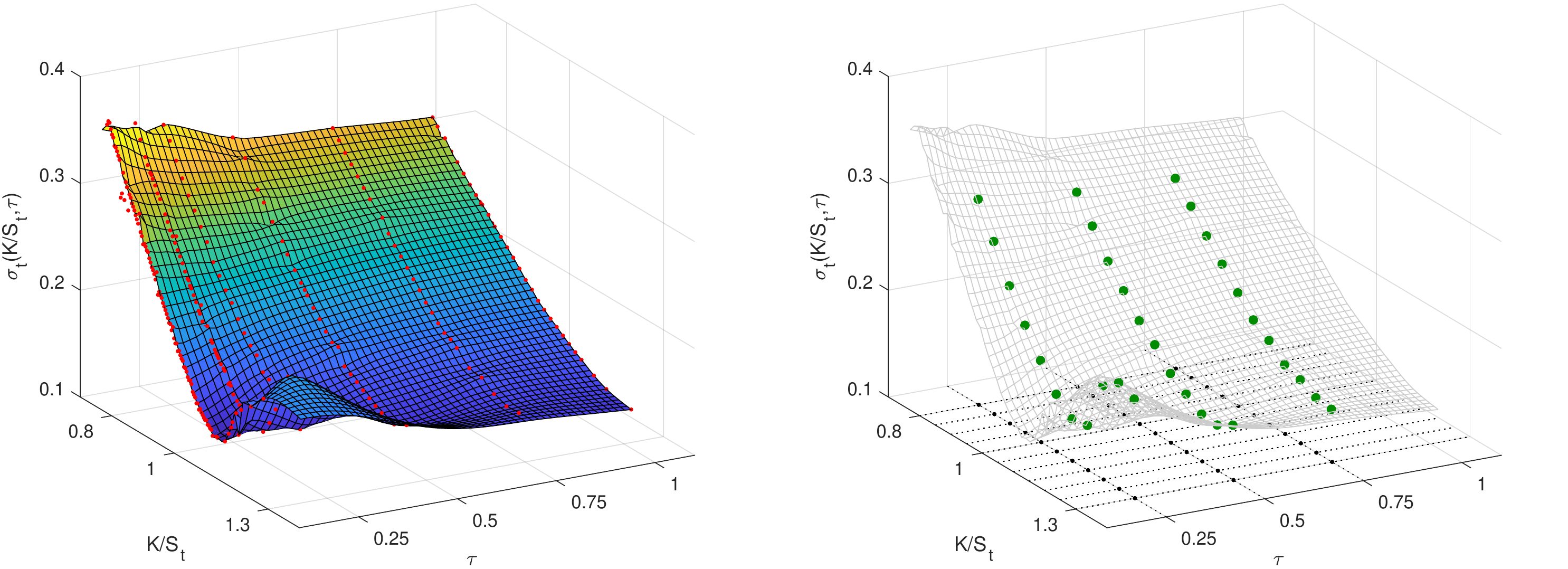}
\caption{IVSF construction and smiles extraction (29-Dec-2010 at 11:00:00; the value of the underlying price is 1,260.89). In the left panel, IVs obtained from market prices (red dots) are used to fit the IVSF. The right panel illustrates the three selected maturities ($\tau$ expressed in years) and the adopted moneyness range and bins (black dots on the moneyness-maturity plane) used in the analyses to slice the surface and extract the sampled smiles  (green dots).}
\label{FIG:surface}
\end{figure}

Although we examine the dynamics of IV smiles for the given maturities and not the IV surface per se, we need to construct the surface first to interpolate the smiles because option quotes are not always available for the exact maturities (i.e., 3, 6, and 9 months) for each minute in the markets. Also, IVs for at-the-money options with a $K/S$ equal to exactly 1 are not typically available. Consequently, we first fit and smooth the IV surfaces, from which we extract smiles for the given maturities and moneyness ranges. The Black-Scholes IV is computed for each available market price, specifically the mid-price, to the maturity-strike domain for every minute on every trading day between 09:31:00 and 16:15:00 from 2006 to 2010.  By using put-call parity, we solve the dividend yield so that the resulting IVs are consistent between the put and call options. The use of out-of-the-money (OTM) and at-the-money (ATM) puts and calls (and the exclusion of the in-the-money options) is motivated by the fact that the OTM and ATM options are those of the most interest as they are traded the most and thus, are the most liquid. This is consistent with VIX, which is calculated by excluding in-the-money options, too. We further obtain the IV surfaces via thin plate spline interpolation \citep{wahba1990spline} and select specific slices of time to maturity for fixed moneyness ranges. The left plot in Figure \ref{FIG:surface} illustrates this procedure at a specific time point, where the surface construction by interpolation is applied to the IVs computed from market mid-prices, while the right plot illustrates the selection of the moneyness range and bins for three specific times to maturity (three, six, and nine months) and the corresponding smiles extracted from the fitted surface. 

A necessary transformation to impose on the data to facilitate the analyses is shifting from the absolute coordinates of strike price $K$ to the relative coordinates of moneyness $m_t = K/S_t$, where $S_t$ denotes the spot price of the underlying price at time $t$. This stems from the variability of $m_t$ as $S_t$ fluctuates. The moneyness range where the dynamics is analyzed is restricted to the interval $m \in \left[ 0.8, 1.3 \right] $. This choice is a trade-off between two competing goals. First, we would like to be able to study a wide range of moneyness to make conclusions about the IV dynamics of deep-out-the-money calls ($m \gg 1$ for OTM calls), as well as deep-out-of-the-money puts ($m \ll 1$ for OTM puts). Second, the width of the moneyness range is restricted by low liquidity at extreme moneyness values \citep{cont2002} and numerical issues related to the surface smoothing. For each maturity (three, six, and nine months), we sample the IV smile in the moneyness bins of width 0.05. The IV observations for the bins (green dots in the right plot of Figure \ref{FIG:surface}) are obtained by averaging the IVs over a finer moneyness grid of width 0.01 within the bin.

Data for implied volatilities were inspected for intra-day seasonality: We found that, on average, the IV is considerably higher during the first minutes of the trading day. To remove this effect, the means of the variables for each observation minute across all trading days in the data sample were subtracted from each observation of the corresponding minute.

\subsection{Jumps in the underlying price data}\label{SUBSEC:Jumps}

Jumps in stock prices are large price movements that cannot be explained statistically by Brownian motion; a jump component is needed to capture such movements. Economically, jumps are related to public or private information arrivals in stock markets \citep{lee2011jumps}, which are associated with liquidity shocks \citep{siikanen2017limit,siikanen2017drives}. The method for detecting jumps in underlying prices used in this research is that of \citep{lee2008jumps}: a non-parametric test applicable to a wide number of financial time series provided that high-frequency data are available. In the present study, this method is used to detect jumps in the underlying price (which is available for every minute). In \citep{lee2008jumps}, the test statistic is based on returns that are scaled by the realized bipower variation. The test statistics are shown to be approximately normally distributed when the underlying log-price comes from a standard Brownian motion ($H_0$), which does not allow for jumps. For a given statistical significance level, a threshold must be exceeded to reject the null hypothesis (in which case, it is unlikely that the observed return comes from a pure jump-free Brownian motion).  For more information about the jump detection method, we refer to \citep{lee2008jumps}.

To answer our research question, we need three data samples of IVs: (i) IV data with a positive single jump in underlying prices,  (ii) IV data with a negative single jump in underlying prices, and (iii) IV data from a period with zero jumps in prices, which serves as reference data. Therefore, as a preliminary step, we label the 1,259 days of data for the S\&P 500 index according to the presence of jumps in the underlying  price. The jump detection implemented with a detection window of one minute leads to a total of 1,226 jumps (both positive and negative, across all the trading hours). Jumps are concentrated in the first hour of the trading day so that even  $\sim 83\%$ occur overnight or during the first trading hour. For this reason, this paper opts to ignore jumps that are detected after 10:30:00 and study the dynamics of smiles only during the course of the morning until 10:30:00 (while including over-night jumps). Moreover, focusing on the first trading hour only addresses the issues regarding the intra-day periodicity of volatility \citep[see e.g.,][]{andersen1997intraday}. Importantly, because we are examining whether single jumps in underlying prices are related to gradual changes in IV, mornings with multiple jumps are excluded.  This leads to two subsamples consisting of 297 mornings with a positive jump and 273 mornings with a negative jump detected. The jump-free subsample includes 398 mornings. However, there are missing values in the data, which come from the option data itself (e.g., missing variables) and the non-reliability of some of the fitted surfaces (e.g., due to multiple options with different market prices for the same moneyness and maturity): 
\begin{itemize}
    \item [--] No-jump sample. Since jumps can occur at any time between 9:31 and 10:30, we remove all the days in the non-jump sample with a missing value in the first two hours, leaving 346 mornings in the jump-free sample. 
    \item [--] Jump samples. We analyze the IV dynamics over 5-, 15-, 20-, 30-, and 60-minute windows. The applicability of a given day depends on the length of the window used: Out of the 297 (273) mornings with positive (negative) jumps, there are 268 (247), 268 (244), 268 (244), 265 (242), and 263 (240) mornings under 5-, 15-, 20-, 30-, and 60-minute windows, respectively, with no missing observations.
\end{itemize}

\section{Methods}\label{SEC:Meth}

\subsection{Principal components and intra-day seasonality effects}\label{SC:Characterization}

In this paper, we analyze not only ATM-IV but also IV smiles. Because smiles consist of multiple data points, we extract principal components to reduce the dimensions. \cite{skiadopoulos2000} explore how many factors are needed to model the dynamics of the IVSF and how they can be interpreted. The technique they use to shed light on these questions is the PCA, which can be considered the method of choice in the literature to answer questions related to the dynamic aspects of the IVSF. It is common practice to identify the number and sources of shocks that move, for example, ATM-IV, with principal components analysis \citep{fengler2003}. This is the same approach employed by \cite{skiadopoulos2000}, who use daily data on futures options on the S\&P 500 index and study the dynamics of the IVSF by forming maturity buckets, across which the IV smiles are averaged and then to which PCA is applied. They extract two principal components that are interpreted as a parallel shift of the surface and a Z-shaped twist of the surface. The extracted components explain, on average, 60\% of the variation of the surface. \cite{panigirtzoglou2004} also extract only two principal components in their study of the dynamics of implied probability distributions of option prices. Some authors \citep[e.g.,][]{cont2002, fengler2003}, however, conclude that three factors are needed to capture an adequate amount of the total variation of the IVSF to satisfactorily model its evolution.

Since the literature commonly exploits the standard PCA and due to the nature of our analyses (which address the dynamics of the smile in the presence of jumps in the underlying price), in this research, we adopt the well-known standard PCA, for which a short description is provided below. Suppose that $ \mathbf{X} \in \mathbf{R}^{n, p}$ is a data matrix of $p$ random variables for $n$ observations, $\mathbf{X} = (\mathbf{x}_1, \ldots , \mathbf{x}_p) $. In our case, $\mathbf{x}_i$ are the observations of the IVs in the $i$-th moneyness bin. PCA replaces the set of $p$ correlated and unordered variables with a set of $ k \leq p $ uncorrelated and ordered linear projections $\mathbf{z}_1, \ldots ,\mathbf{z}_k$ of the original variables \citep{izenman2008}. The linear projections can be written as follows:
\begin{equation}\label{EQ:LINPROJ}
\mathbf{z}_j = \mathbf{b}{_j}\mathbf{X}^T = b_{j1}\mathbf{x}_1 + \ldots + b_{jp}\mathbf{x}_p, \qquad j = 1,2, \ldots ,k \nolinebreak ,
\end{equation}
where $ \textbf{b}_j $ is the vector of the loadings for the $j$-th component. The goal is to find the projections that minimize the loss of information. When the coefficient vectors $ \mathbf{b}_j$ are picked so that the projections $z_j$ are ranked in decreasing order of variance and that $z_j$ is uncorrelated with all the $z_i$ (for $ i < j $), we call the linear projections of Eq. (\ref{EQ:LINPROJ}) the $j$-th principal components of $\mathbf{X}$. 

Given the $p \times k$ loading matrix $\textbf{B} = \left( \mathbf{b}_1^T, \ldots, \mathbf{b}_k^T \right)$ of the first $k$ principal components, the elements of the $n	\times k $ matrix $\mathbf{S} = \mathbf{XB}$ representing the data matrix $\mathbf{X}$ on the principal components space, are commonly called \textit{scores}. The $j$-th column of $\mathbf{S}$ collects the scores associated with the $j$-th principal component. For further details and proofs, see, for example, \citep{izenman2008}.

In this study, we extract three principal components and study the dynamics, specifically $\Delta IV_t = IV_t - IV_{t-1}$. It is common practice in the field not to deal with the IV itself but with its changes \citep{skiadopoulos2000,fengler2003,panigirtzoglou2004, cont2002}. We extracted the principal components for $\Delta IV$ using all the data. Table \ref{TAB:PERCVAR} shows the percentages of the variance explained by the first three principal components, and the table illustrates that the loss of information (unexplained variance) when characterizing the smile with these components is quite moderate, which is consistent with \citep{cont2002,fengler2003}. However, to the authors' knowledge, no studies have been conducted in which PCA was applied to study the \textit{intra-day} dynamics of the smile. Previous research only relied on data sampled at daily intervals.

\begin{table}[htbp]
  \centering
  \caption{Interpretation of the first three principal components. ATM-PC stands for the principal component based on at-the-money options, OTM-Call-PC for the principal component based on out-of-the-money calls, and OTM-Put-PC for the principal component based on out-of-the-money puts.}
    \begin{tabular}{cccc}
    Maturity & PC\textsubscript{1} & PC\textsubscript{2} & PC\textsubscript{3} \\
    \midrule
    3 months & OTM-Call-PC & ATM-PC & OTM-Put-PC \\
    6 months & OTM-Call-PC & ATM-PC & OTM-Put-PC \\
    9 months & OTM-Call-PC & OTM-Put-PC & ATM-PC \\
    \bottomrule
    \end{tabular}
  \label{tab:PCInterpretation}%
\end{table}%

\begin{table}[htbp]
  \centering
  \caption{Percentage of the total variance explained by each component. ATM-PC stands for the principal component based on at-the-money options, OTM-Call-PC for the principal component based on out-of-the-money calls, and OTM-Put-PC for the principal component based on out-of-the-money puts.}
    \begin{tabular}{ccccc}
    Maturity & OTM-Call-PC & ATM-PC & OTM-Put-PC & Total \\
    \midrule
    3 months & 66.58\% & 10.60\% & 08.74\% & 91.16\% \\
    6 months & 61.85\% & 15.15\% & 11.50\% & 88.51\% \\
    9 months & 60.48\% & 12.83\% & 15.54\% & 88.86\% \\
    \bottomrule
    \end{tabular}%
  \label{TAB:PERCVAR}%
\end{table}%

\begin{figure}[!h]
\centering
\includegraphics[scale=0.65]{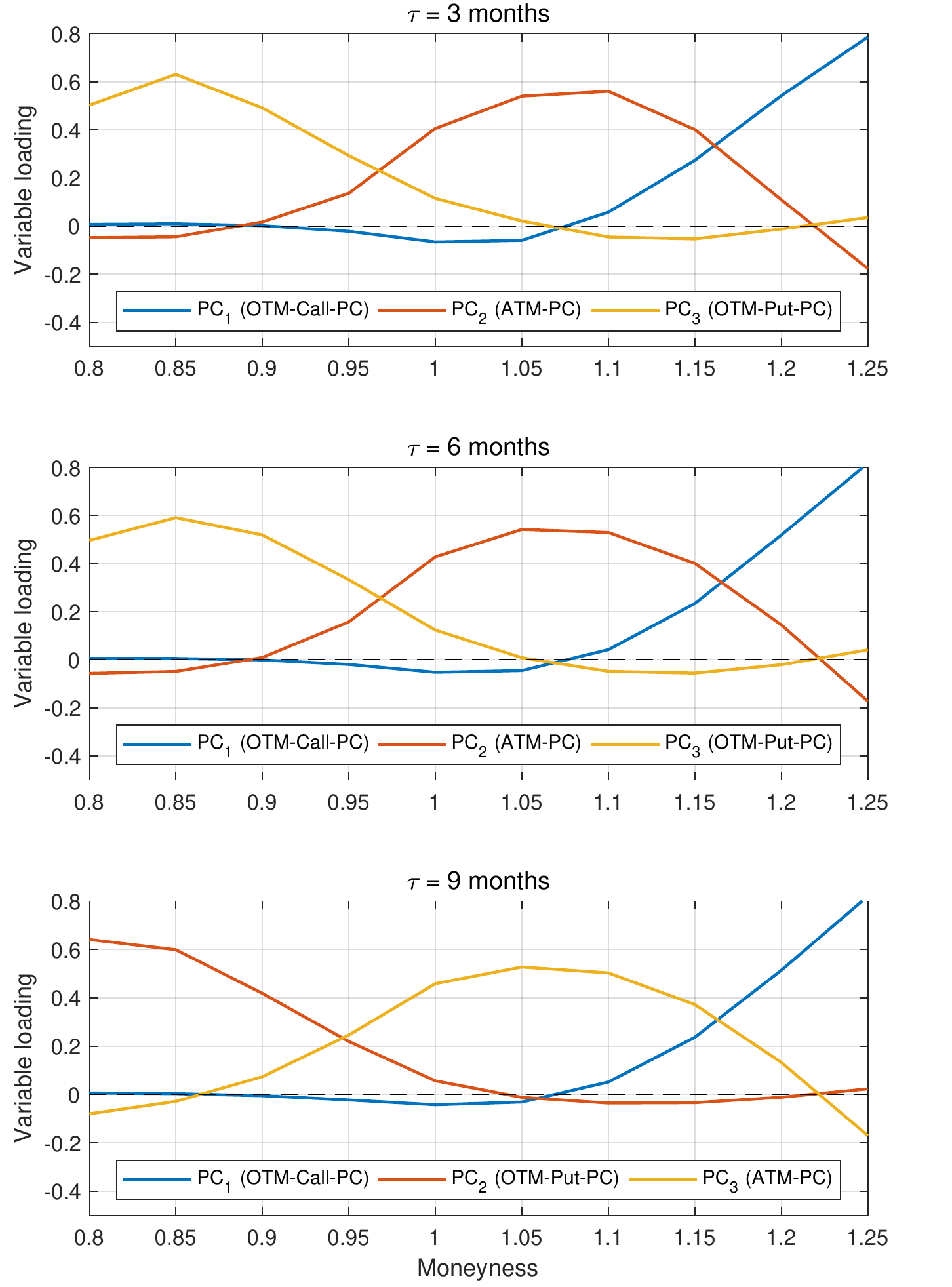}
\caption{Varimax-rotated variable loadings of the first three principal components for differences in IV ($\Delta IV$).} 
\label{fig:loadings}
\end{figure}

The interpretability of the PC we obtain is crucial for making meaningful conclusions about the smile behavior around underlying price jumps. The principal components can be interpreted by inspecting what are commonly called parallel coordinate plots, where the loadings of each principal component are plotted against the indices of the original variables (Figure \ref{fig:loadings}). Loadings indicate how much each original variable contributes to a principal component. Commonly, (Varimax) rotation is applied to the original principal components to facilitate their interpretation while preserving their lack of correlation. This method generally yields to components that have a clear interpretation in terms of the original variables.

Figure \ref{fig:loadings} can be inspected to seek an interpretation of the three components.
For three- and six-month maturities, the first PC (blue curve) is interpreted as representing OTM call options because the bins of moneyness greater than one are highly loaded on this component, while the other two components exhibit lower loadings in the same range. The second PC (red curve) is highly loaded in the moneyness bins corresponding to the out-of-the-money put option. Similarly, the third PC (yellow curve) is interpretable as ATM (put and call) options. Moving to the nine-month maturity, we notice that the second component corresponds to OTM puts and the third to ATM options, while the interpretation of the first component remains unchanged. 
Based on Figure \ref{fig:loadings}, the principal components are renamed according to their interpretations. Table \ref{tab:PCInterpretation} provides an overview of their interpretation across the three maturities.

As in the case of ATM-$\Delta IV$, the principal components are also considerably higher during the first minutes of the trading day. To remove the seasonality effect, we subtract the minute-by-minute mean of the scores under no jumps from the whole data (both under jumps and not) in the corresponding minute. Involving the scores under jumps in the computation of the mean would clearly bias the seasonality profile, thus their exclusion. The same procedure is applied to ATM-$\Delta IV$.

\subsection{Regression model\label{SEC:RegMod}}

Methodologically, we suggest event-based research on implied volatility. Instead of investigating how cumulative abnormal returns behave after given news events, which is a typical setup in the event study literature \citep{binder1998event}, we analyze cumulative changes in implied volatility after positive and negative return jumps compared to days where no return jumps are detected. As mentioned earlier, we focus on the first trading hours as 83\% of jumps occur before 10:30. 

Using the following regression model, which includes indicators for the jump direction, we analyze total and delayed reactions in the IV:
\begin{equation}\label{EQ:k0}
\overline{\Delta IV} = \beta_0 + \beta_P I_P +\beta_N I_N +\beta_{IV} \overline{IV},
\end{equation}
where $\overline{\Delta IV}$ is the cumulative IV over a given time window  (5, 15, 20, 30, or 60 minutes); $I_P$  and $I_N$ are indicator variables; $I_P=1$ if there is a positive jump on a given day (morning); $I_N=1$ if there is a negative jump, and if there are no jumps, then $I_P = I_N = 0$.  Additionally, the regression is controlled by the level of the IV because of the heteroscedasticity in volatility, which is a well-known financial fact:  the variance in volatility depends on the volatility level. The level of volatility, $\overline{IV}$, is constructed by averaging the minute-by-minute IVs for the ATM options for a given maturity over the first 60 minutes on this day (independently, if out or at-the-money IV is studied). 

Parameter estimates of $\beta_P$  and $\beta_N$ show the {\em total impact} of positive and negative jumps on the IV, respectively, if we cumulate $\Delta IV$ from the {\em beginning} of the return jump period onwards (i.e., we do {\em not exclude the first reaction} in the IV to the calculation of $\overline{\Delta IV}$). To study the existence of {\em gradual movements} in the implied volatility, the null hypothesis about efficient markets around an underlying price jump is verified by {\em excluding} the first movement in the IV, which is simultaneous with a return jump, from the formation of $\overline{\Delta IV}$. That is, in this case, $\Delta IV$ is cumulated from the {\em end} of the return jump period onwards so that IV's first reaction, which is located within the jump period, is not included. If the parameter estimate of $\beta_P$ is still negative and statistically significant ($\beta_N$ is still positive and statistically significant), then we reject the null hypothesis and conclude that there is a negative (positive) post-drift in the IV after a positive (negative) return jump. We run the regression separately, including and excluding the first IV movements and separately for each IV variable (i.e., ATM-IV and three principal components) to analyze whether the IV smile keeps its shape while reacting to return jumps. Our main results consider the post-windows of 5 and 30 minutes, and the results for windows of 15, 20 and 30 minutes are presented in the Appendix.

The regression model requires three data samples of the IVs: (i) IV data with a positive single jump in underlying prices,  (ii) IV data with a negative single jump in underlying prices, and (iii) IV data on a period with zero jumps in prices, which serves as reference data. On a given day in data samples (i--ii), the IV is observed from the beginning of the detected return jump period onwards, but the construction of data sample (iii) is not that straightforward because, by construction, no return jump periods are detected, and therefore, there are no natural starting points for the observation windows. We solve this problem by sampling the time intervals for the reference sample using the empirical distribution of the time stamps of the detected positive and negative jump periods. Hence, the time-stamp that defines the starting point of the reference data observation window on a given day is randomly selected according to the empirical distribution of the return jump timestamps, so the location of the data observation window is equivalently distributed between no-jump-days and jump-days.

\section{Results}\label{SEC:Results}

Table \ref{TAB:IV_Regression_Results_Inc_First_Reaction} shows the regression results 
by cumulating minute-by-minute changes in the IV over (a) 5 minutes and (b) 60 minutes. To examine the overall impact of a return jump on the IV, in this table, the movements in the IV observed from the return jump window are not excluded. That is, reactions in the IV that are simultaneous with the return jumps are included in the cumulative $\Delta IV$ to see the total magnitude and direction of the overall IV reaction. In the table, the regression results are available not only for at-the-money IV but also principal components that capture the volatility smile for 3-, 6-, and 9- month maturities. The parameter estimates for $\beta_p$ are evidentially negative and, correspondingly, the parameter estimates of $\beta_n$ are positive. This means that, on average, a positive (negative) return jump drives the IV down (up), which is in line with the existing literature \citep{eraker2004stock,jacod2010price,todorov2011volatility,bandi2016price}. The results are consistent across ATM-IV and all the principal components, and therefore, the whole IV smile moves in the opposite direction compared to the return jump direction, not only at-the-money implied volatility.  

\begin{sidewaystable}
\caption{Regression results {\em including} the IV moments observed from the return jump window. Data is observed from 09:30 to 10:30, and return jumps are detected using 1-minute windows with the significance level of 1\%. }\label{TAB:IV_Regression_Results_Inc_First_Reaction}
\centering
\resizebox{0.9\textheight}{!}
{
\begin{tabular}{lcccc|cccc}
      & \multicolumn{4}{c|}{\textbf{5 minutes}} & \multicolumn{4}{c}{\textbf{60 minutes}} \\
      & \boldmath{}\textbf{$\beta_0$}\unboldmath{} & \boldmath{}\textbf{$\beta_p$}\unboldmath{} & \boldmath{}\textbf{$\beta_n$}\unboldmath{} & \boldmath{}\textbf{$\beta_{IV}$}\unboldmath{} & \boldmath{}\textbf{$\beta_0$}\unboldmath{} & \boldmath{}\textbf{$\beta_p$}\unboldmath{} & \boldmath{}\textbf{$\beta_n$}\unboldmath{} & \boldmath{}\textbf{$\beta_{IV}$}\unboldmath{} \\
\hline
\textbf{ATM-IV} &       &       &       &       &       &       &       &  \\
3 months & 0.082 & -0.440 & 0.471 & -2.775E-03 & 0.208 & -0.352 & 0.565 & -9.943E-03 \\
      & (0.068*) & (6.454E-22***) & (4.018E-23***) & (0.141) & (3.267E-04***) & (1.800E-09***) & (4.044E-20***) & (4.217E-05***) \\
6 months & 0.057 & -0.275 & 0.356 & -2.853E-03 & 0.067 & -0.230 & 0.422 & -3.760E-03 \\
      & (0.074*) & (2.866E-21***) & (2.156E-31***) & (0.038*) & (0.157) & (8.751E-08***) & (1.177E-20***) & (0.062*) \\
9 months & 0.022 & -0.215 & 0.305 & -1.151E-03 & 0.030 & -0.214 & 0.335 & -1.484E-03 \\
      & (0.396) & (6.901E-23***) & (1.247E-39***) & (0.294) & (0.447) & (1.305E-10***) & (1.116E-21***) & (0.374) \\
\textbf{ATM-PC} &       &       &       &       &       &       &       &  \\
3 months & 0.016 & -0.785 & 0.821 & 8.217E-04 & 0.115 & -0.725 & 0.936 & -7.287E-03 \\
      & (0.834) & (3.300E-22***) & (1.970E-22***) & (0.801) & (0.303) & (2.161E-10***) & (3.909E-15***) & (0.116) \\
6 months & 0.010 & -0.603 & 0.627 & 9.267E-04 & 0.086 & -0.532 & 0.700 & -3.349E-03 \\
      & (0.866) & (6.335E-28***) & (5.755E-28***) & (0.711) & (0.317) & (1.532E-11***) & (2.195E-17***) & (0.374) \\
9 months & -4.427E-03 & -0.452 & 0.574 & 6.638E-04 & 0.061 & -0.447 & 0.640 & -2.802E-03 \\
      & (0.928) & (5.771E-26***) & (5.325E-37***) & (0.753) & (0.427) & (8.035E-12***) & (9.896E-21***) & (0.396) \\
\textbf{OTM-Call-PC} &       &       &       &       &       &       &       &  \\
3 months & -0.058 & -0.943 & 0.821 & 7.954E-03 & 0.337 & -0.646 & 0.888 & -0.014 \\
      & (0.803) & (6.549E-05***) & (8.210E-04***) & (0.422) & (0.205) & (0.016*) & (1.472E-03**) & (0.206) \\
6 months & -0.321 & -0.559 & 0.636 & 0.014 & -0.047 & -0.369 & 0.771 & -1.915E-03 \\
      & (0.016*) & (3.675E-06***) & (4.287E-07***) & (0.012*) & (0.750) & (6.419E-03**) & (4.671E-08***) & (0.761) \\
9 months & -0.082 & -0.484 & 0.417 & 5.467E-03 & 9.392E-03 & -0.337 & 0.527 & -3.001E-03 \\
      & (0.306) & (1.848E-12***) & (4.356E-09***) & (0.111) & (0.924) & (5.851E-05***) & (1.605E-09***) & (0.478) \\
\textbf{OTM-Put-PC} &       &       &       &       &       &       &       &  \\
3 months & 0.137 & -0.803 & 0.972 & -5.886E-03 & 0.247 & -0.725 & 0.946 & -8.752E-03 \\
      & (0.143) & (2.673E-17***) & (1.089E-22***) & (0.130) & (0.030*) & (6.522E-10***) & (1.238E-14***) & (0.074*) \\
6 months & 0.064 & -0.560 & 0.671 & -2.359E-03 & 0.039 & -0.487 & 0.757 & -2.171E-03 \\
      & (0.325) & (1.052E-20***) & (1.215E-26***) & (0.393) & (0.677) & (5.859E-09***) & (5.513E-18***) & (0.575) \\
9 months & 0.093 & -0.414 & 0.540 & -3.763E-03 & 0.065 & -0.385 & 0.630 & -4.242E-03 \\
      & (0.042*) & (7.770E-26***) & (1.272E-38***) & (0.056*) & (0.372) & (3.263E-10***) & (1.410E-22***) & (0.171) \\
\end{tabular}%
}
\end{sidewaystable}

Secondly, and more interestingly, to analyze whether there are post-movements in the implied volatility, we exclude the IV movements observed from the 1-minute return jump window (i.e., we exclude the simultaneous reactions in the IV with return jumps). Table \ref{TAB:IV_Regression_Results_Excl_First_Reaction} shows the parameter estimates under these settings. Again, the results are available for at-the-money IV and three principal components for 3-, 6-, and 9-month maturities and (a) 5 and (b) 60-minute windows over which the $\Delta IV$ is cumulated. Here, the observations of IV changes from the first minute are now excluded, and therefore, in fact, results from five-minute periods are based on four observations that occur after the 1-minute return jump window. Correspondingly, results from 60-minute periods are based on 59 minute-by-minute observations. Results for 15-, 20-, and 30-minute windows are available in Table \ref{TAB:IV_Regression_Results_Exc_First_Reaction_5_60} in the Appendix.

Panel (a) of Table \ref{TAB:IV_Regression_Results_Excl_First_Reaction} provides evidence that there are post-reactions within the four minutes that follow the one-minute return jump window. These results hold for ATM-IV, ATM-PC, and OTM-Put-PC, for which the parameter estimates of $\beta_p$ and $\beta_n$ are negative and positive, respectively, and hence, we confirm that the IV continues to decrease (increase) after a positive (negative) return jump over the following four minutes. However, OTM-Call-PC is an exception; there are no post-reactions identified for it over the post-jump four-minute period. Therefore, the post dynamics of the IV smile is not consistent across the moneyness: the IV based on at-the-money options and out-of-the-money put options (and correspondingly, by put-call parity, in-the-money call options) react to individual return jumps gradually, while the IV based on out-of-the-money call options (and correspondingly, by put-call parity, in-the-money put options) immediately converges to its new level within the return jump period. Therefore, the implied volatility smile is adjusted to jumps in underlying's return asymmetrically. Moreover, across all the maturities and moneyness groups, the parameter estimates are of higher magnitude for $\beta_n$ compared to $\beta_p$, meaning that the post-reactions in the IV are stronger after negative rather than positive return jumps. 

We also ran the analysis over a 60-minute period to determine whether post-reactions in the IV gain or vanish within a 60-minute window. We find that the parameter estimates for $\beta_n$ are of an even greater magnitude, and estimates based on the out-of-the-money call principal component become significant for 6- and 9-month options. However,  $\beta_p$ becomes less significant in almost every case. Interestingly, for the out-of-the-money call principal component, the estimate of $\beta_p$ is {\em positive} and significant, meaning that there is a significant correction effect after the first (simultaneous) IV movement. Overall, gradual IV movements persist---or even gain---for negative return jumps, while they are revised---or even vanish---for positive return jumps. 

\begin{sidewaystable}
\caption{Regression results {\em excluding} the IV moments observed from the return jump window. Data is observed from 09:30 to 10:30 and return jumps are detected using 1-minute windows with the significance level of 1\%.}\label{TAB:IV_Regression_Results_Excl_First_Reaction}
\centering
\resizebox{0.9\textheight}{!}
{
\begin{tabular}{lcccc|cccc}
      & \multicolumn{4}{c|}{\textbf{5 minutes}} & \multicolumn{4}{c}{\textbf{60 minutes}} \\
      & \boldmath{}\textbf{$\beta_0$}\unboldmath{} & \boldmath{}\textbf{$\beta_p$}\unboldmath{} & \boldmath{}\textbf{$\beta_n$}\unboldmath{} & \boldmath{}\textbf{$\beta_{IV}$}\unboldmath{} & \boldmath{}\textbf{$\beta_0$}\unboldmath{} & \boldmath{}\textbf{$\beta_p$}\unboldmath{} & \boldmath{}\textbf{$\beta_n$}\unboldmath{} & \boldmath{}\textbf{$\beta_{IV}$}\unboldmath{} \\
\hline
\textbf{ATM-IV} &       &       &       &       &       &       &       &  \\
3 months & -1.570E-04 & -0.096 & 0.136 & 7.952E-04 & 0.125 & -0.019 & 0.218 & -5.759E-03 \\
      & (0.996) & (1.123E-03**) & (9.822E-06***) & (0.523) & (0.020*) & (0.720) & (1.055E-04***) & (0.011*) \\
6 months & -2.111E-03 & -0.055 & 0.080 & -2.323E-05 & 0.038 & -7.963E-03 & 0.149 & -2.434E-03 \\
      & (0.915) & (1.770E-03**) & (1.168E-05***) & (0.978) & (0.323) & (0.820) & (4.081E-05***) & (0.139) \\
9 months & 0.015 & -0.044 & 0.062 & -5.027E-04 & 0.042 & -0.028 & 0.109 & -2.367E-03 \\
      & (0.284) & (1.834E-04***) & (3.843E-07***) & (0.405) & (0.194) & (0.308) & (1.319E-04***) & (0.086*) \\
\textbf{ATM-PC} &       &       &       &       &       &       &       &  \\
3 months & -0.063 & -0.092 & 0.270 & 2.235E-03 & 0.046 & -0.048 & 0.367 & -5.490E-03 \\
      & (0.173) & (0.050*) & (3.722E-08***) & (0.246) & (0.631) & (0.618) & (2.634E-04***) & (0.165) \\
6 months & 0.028 & -0.115 & 0.160 & -7.791E-04 & 0.116 & -0.049 & 0.227 & -5.304E-03 \\
      & (0.363) & (3.862E-05***) & (3.489E-08***) & (0.549) & (0.103) & (0.451) & (7.074E-04***) & (0.088*) \\
9 months & -9.982E-03 & -0.097 & 0.104 & 5.960E-04 & 0.036 & -0.073 & 0.186 & -2.785E-03 \\
      & (0.701) & (1.474E-05***) & (7.651E-06***) & (0.596) & (0.577) & (0.181) & (1.008E-03**) & (0.317) \\
\textbf{OTM-Call-PC} &       &       &       &       &       &       &       &  \\
3 months & 0.135 & 0.061 & 0.171 & -6.746E-03 & 0.431 & 0.459 & 0.338 & -0.029 \\
      & (0.282) & (0.630) & (0.196) & (0.207) & (0.052*) & (0.041*) & (0.146) & (2.069E-03**) \\
6 months & -0.019 & -0.058 & 0.055 & 1.501E-03 & 0.208 & 0.161 & 0.218 & -0.014 \\
      & (0.758) & (0.311) & (0.348) & (0.578) & (0.081*) & (0.139) & (0.053*) & (5.936E-03**) \\
9 months & 3.650E-03 & -0.017 & 0.094 & -1.425E-03 & 0.217 & 0.073 & 0.149 & -0.013 \\
      & (0.939) & (0.679) & (0.024*) & (0.482) & (8.204E-03**) & (0.290) & (0.039*) & (2.692E-04***) \\
\textbf{OTM-Put-PC} &       &       &       &       &       &       &       &  \\
3 months & 0.089 & -0.148 & 0.176 & -2.630E-03 & 0.124 & 0.026 & 0.248 & -6.473E-03 \\
      & (0.106) & (7.330E-03**) & (2.163E-03**) & (0.253) & (0.211) & (0.794) & (0.018*) & (0.128) \\
6 months & -0.019 & -0.127 & 0.108 & 1.161E-03 & -0.018 & -0.068 & 0.180 & 8.416E-04 \\
      & (0.599) & (1.252E-04***) & (1.579E-03**) & (0.453) & (0.813) & (0.311) & (9.897E-03**) & (0.788) \\
9 months & 0.053 & -0.055 & 0.145 & -2.278E-03 & 0.092 & -0.017 & 0.247 & -6.243E-03 \\
      & (0.028*) & (5.787E-03**) & (5.708E-12***) & (0.027*) & (0.133) & (0.739) & (4.223E-06***) & (0.018*) \\
\end{tabular}%

}
\end{sidewaystable}

\begin{figure}
\centering
\includegraphics[scale=0.65]{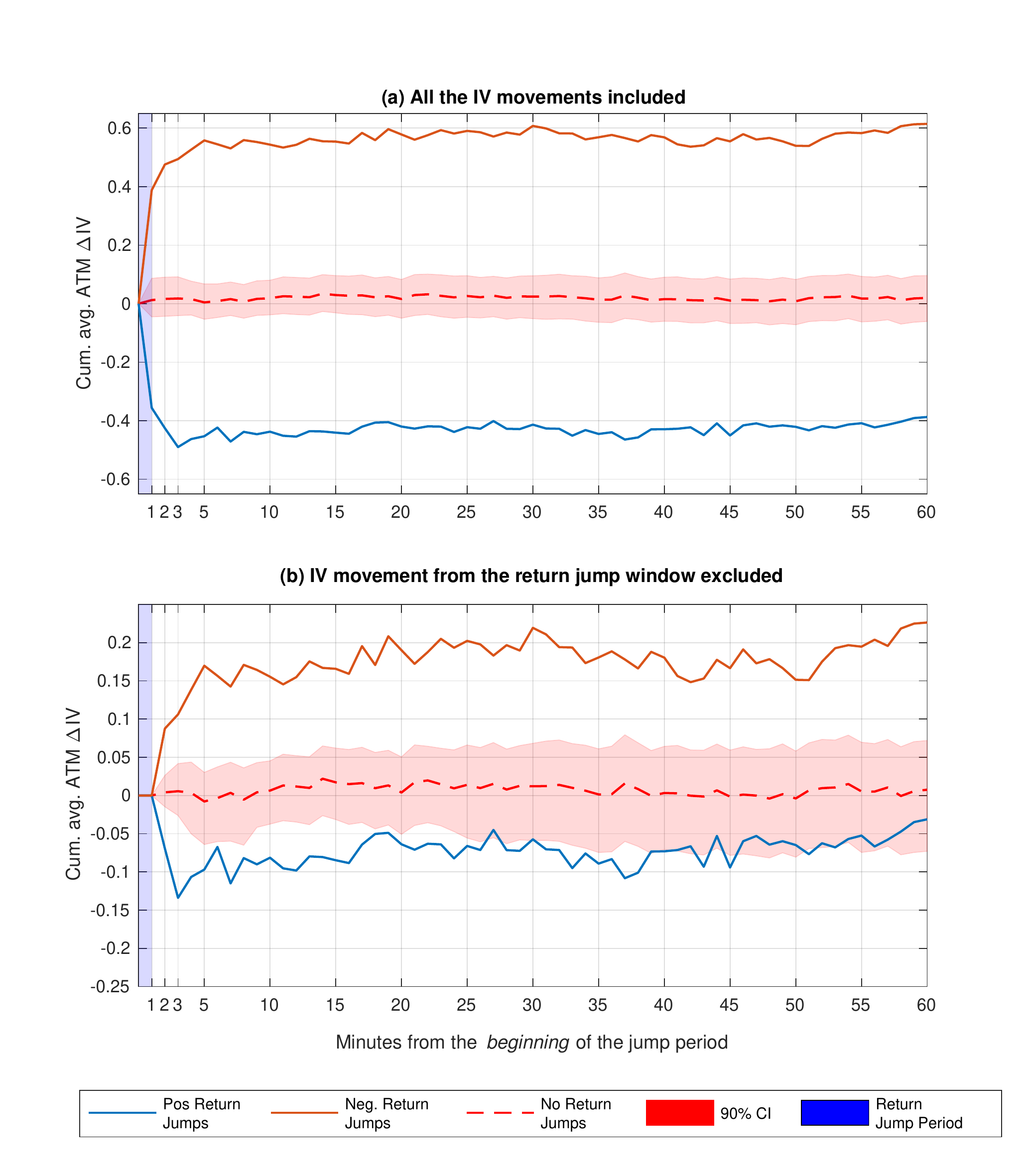}
\caption{Average cumulative movements in implied volatility (IV) after the arrival of return jumps. Here, the IV is measured by {\em at-the-the-money IV}. Time index refers to minutes from the beginning of the 1-minute return jump period. That is, the return jump period is between 0 and 1 minutes (blue interval), during which jumps have arrived, and the post-movements in the IV are from minute 1 onward. Cumulative IV movements after positive jumps are in blue, and after negative jumps are in red. The dashed curve represents cumulative IV movements when no return jumps have occurred. Panel (a) plots the cumulative IV, including the IV movement over the return jump period. Panel (b) excludes the simultaneous IV movement with the return jump. The 90\% confidence interval is obtained by numerical bootstrapping.} 
\label{FIG:IV-Dynamics_ATM_90}
\end{figure}

The IV dynamics after return jumps is demonstrated in Figure \ref{FIG:IV-Dynamics_ATM_90}, where we plot the average cumulative changes in the at-the-money IV over 60 minutes from the beginning of the return jump interval. In the figure, the IV is based on three-month at-the-money options, and the IV is expressed as a percentage. To differentiate the response of IV to positive versus negative changes in the underlying asset price, the average IV trajectories for positive and negative return jumps are plotted separately by blue and red curves, respectively, and periods in which no return jumps are detected are represented by a dashed line. The 90\% confidence interval is calculated by bootstrapping the distribution of the observations from no-jump periods after randomly initialized starting points, as outlined in Section \ref{SEC:RegMod}. The bootstrapping procedure involves 7000 draws for each minute interval. Panel (a) plots the average cumulative IV {\em including} the movement from the return-jump time-interval (from minute 0 to minute 1). That is, in Panel (a), the first movement, which is observed from the beginning to the end of the one-minute jump detection period, represents the average IV movement that is simultaneous with the return jump. In contrast, these IV movements that are simultaneous with return jumps are excluded in Panel (b). Although the deseasonalization of the IV variables sets the adjusted variables to zero in the reference sample, the mean values of the reference data in the plots slightly deviate from zero due to the randomized locations of the data windows (starting points of the windows are drawn from the empirical distribution, see Section \ref{SEC:RegMod}).

A number of observations can be made from the figure. In Panel (a), the first (simultaneous) movement in the IV is in the opposite direction compared to the return jumps, which was also confirmed in Table \ref{TAB:IV_Regression_Results_Inc_First_Reaction}. Moreover, the first movement is considerably large compared to the later ones. However, Panel (b) shows that even if the first movements are excluded, the remaining movements, on average, drive the IV to the outside of the confidence interval (note that the two sub-figures are with different y-scale). In the case of negative return jumps, the IV gets its largest movements over the first five minutes on average but continues to grow over the next 20 minutes, which is a considerably long time. However, the IV dynamics after positive return jumps is quite different. Panel (b) shows that the two first post-movements in the IV are considerably high, but then, after the third minute, the IV stops to decrease and no additional directed gradual changes can be observed. In fact, one could say that there is a small 'correction' in the IV between the third and sixth minutes. The average IVs at the 60th minute show the asymmetry between IV dynamics after positive and negative return jumps clearly: while the IV has drifted upwards and remained outside of the confidence interval over 60 minutes after negative return jumps, the first post-reactions in IV, which are considerably strong, have almost vanished within an hour after positive return jumps. This is exactly what was observed in Table \ref{TAB:IV_Regression_Results_Excl_First_Reaction}. Overall, there are clear post-return-jump reactions in the IV over a couple of minutes for both positive and negative return jumps, but longer-term dynamics over the next hour is quite asymmetric.

Figures \ref{FIG:IV-Dynamics_ATM_PC_90}--\ref{FIG:IV-Dynamics_OTM-Put_PC_90} placed in Appendix show the corresponding average cumulative movements for the three principal components based on 90-day options (figures for the other maturities are available upon request). Clearly, Figure \ref{FIG:IV-Dynamics_ATM_PC_90} on the ATM-PC is close to Figure \ref{FIG:IV-Dynamics_ATM_90} on ATM-$\Delta IV$ as they both represent at-the-money options. Figure \ref{FIG:IV-Dynamics_OTM_Call_PC_90} represents the cumulative IV for out-of-the-money call (in-the-money put) options. As mentioned earlier, the estimates of $\beta_p$ for OTM-Call-PC in Table \ref{TAB:IV_Regression_Results_Excl_First_Reaction} suggest that the post-jump IV movements correct the first negative reaction. Indeed, this correction effect is observable in Figure \ref{FIG:IV-Dynamics_OTM_Call_PC_90}: The first IV movement, which is simultaneous with a positive return jump, is negative, and then the later IV movements revise the IV level up. If we exclude the first reaction (Panel B), then we can see that the IV is drifting upward to a point where it is outside of the confidence interval. Finally, Figure \ref{FIG:IV-Dynamics_OTM-Put_PC_90} for OTM-Put-PC is quite similar to the figures on at-the-money options.

\section{Robustness Checks}
We use three approaches to check the robustness of our results. First, instead of detecting return jumps at the 1\% significance level, we apply the 5\% level. These results are reported in Table \ref{TAB:IV_Regression_Results_Exc_First_Reaction_5_60_5Pros} in the Appendix. The results are quite similar, and the conclusions remain the same. Secondly, in Table \ref{TAB:IV_Regression_Results_Exc_First_Reaction_5_60_1230} in the Appendix, results are reported from an alternative analysis where the focus is not only in the first trading hour between 09:30 and 10:30, but until 12:30 (i.e., data from the first three trading hours are used). All the other settings are the same as in the baseline analysis (detecting return jumps at 1\% significance level using 1-minute detection periods). Again, the results are very similar to the main analysis, and the conclusions remain the same. Finally, we run the regression 1,000 times with different realizations of the reference data. In particular, because the reference data observation windows are randomly located according to the empirical distribution of the return jump timestamps, the reference dataset is different for each regression, and hence, the regression results differ (see Section \ref{SEC:RegMod}). Otherwise the regressions follow the baseline: they exclude the IV moments observed from the return jump window, data is observed from 09:30 to 10:30, and return jumps are detected using 1-minute windows with a significance level of 1\%. Table \ref{TAB:IV_Regression_Results_Exc_First_Reaction_5_60_1000Reg} reports the results for 1,000 regressions with different realizations of reference data. In the table, the mean, standard deviation, and 2.5\% and 97.5\% quantiles (Q2.5\% and Q97.5\%) are reported for the parameter estimates and p-values. Panel (a) reports results for the 5-minute window and Panel (b) for the 60 minute IV window. This robustness analysis shows that the observed results are not random but systematic over 1,000 different realizations of the reference datasets, which confirms our analysis above.

\section{Economic Significance}
Table \ref{TAB:IV_Regression_Results_Excl_First_Reaction} reports that $\beta_n = 0.218$ for 3-month at-the-money implied volatility at the 60-minute post window. That is, compared to the situation where there have been no jumps, the implied volatility has grown on average 0.218 more {\em after} the arrival of a negative jump over the next 60 minutes. Is this economically significant? 

For the at-the-money option, the relationship between call price and implied volatility can be approximated \citep{brenner1988simple} as
\[
c_{BS}\left(\sigma_{t,T}^{ATMBS}; t\right) = S(t) \sigma_{t,T}^{ATMBS} \sqrt{\frac{T-t}{2\pi}},
\]
where $\sigma_{t,T}^{ATMBS}$ is the ATM-IV with maturity $T-t$. Therefore, if the at-the-money implied volatility has increased by $\Delta \sigma_{t,T}^{ATMBS}$, ceteris paribus, the return for a long call option over 60 minutes would be $\Delta \sigma_{t,T}^{ATMBS} / \sigma_{t,T}^{ATMBS}$. Given that, after the negative return jump, the ATM-IV would be at 15\% or 20\% and that it increases by 0.218\% to 15.218\% or 20.218\%, the corresponding option return would be 1.54\% or 1.09\%. This would represent a good return for a 60-minute period, but abnormal returns in option markets are not possible because of the transaction costs. We examined the realized profits on long (short) call option positions that took place at the end of the negative (positive) return jump interval and lasted 5 or 60 minutes. The realized option returns on mid-prices were positive on average, but when taking the option spread into account, the net returns were clearly negative on average. Also, the realized net-returns for put-options  were negative on average. Overall, we provide evidence that option markets do not incorporate new information about the expectations of option payoffs immediately, which is contrary to the efficient market hypothesis. At the same time,  observed informational inefficiency does not imply abnormal returns. 

\section{Conclusion and Discussion}\label{SEC:Conclusions}

This research provides evidence that option markets adjust the implied volatility (IV) of the S\&P 500 index options gradually after the arrival of a single jump in the underlying return. This phenomena regarding delayed reactions in the IV is particularly significant for at-the-money options and out-of-the-money puts, while the IV of out-of-the-money call options seems to converge to its new level simultaneously (within the same minute) with the return jump. Therefore, out-of-the-money calls seem to be priced more efficiently than out-of-the-money puts or at-the-money options. 

For at-the-money options and out-of-the-money puts, we find that that the IV gradually and clearly increases over a few minutes that follow a {\em negative} return jump, and the gradually driven IV persists over the next trading hour after negative return jumps. In zero-transaction markets, option traders could benefit on average from having a long position on at-the-money options right after the arrival of negative return jumps and terminate the option position after 30 minutes when the IV has converged to its new level. However, transaction costs (option price spread) make the options' net-returns negative, and thus, abnormal returns are not observed. Regarding {\em positive} return jumps, while the IV gradually decreases within the next few minutes, the gradual IV movements are not long lasting as they vanish over the following 60 minutes. Therefore, in the case of positive return jumps, it is the first (simultaneous) reaction in the IV that matters in the long-term. These results were found to be robust across different research settings.

The inefficiency of option markets calls for carefulness in the calibration of option models to option market data, which is often necessary after important market shocks. Often, the option pricing models (including, for example, stochastic volatility and jumps) are calibrated by minimizing the squared difference between model-based and market-based implied volatilities. However, if markets do not price options correctly, using the market prices of options without alternative data sources can lead to biased parameter estimates. Alternatively, models can be estimated using time-series data on underlying returns\footnote{There are different estimation methods for option pricing models with stochastic volatility, such as maximum likelihood estimation \citep{ai2007maximum}, Markov Chain Monte Carlo methods \citep{eraker2003impact,yang2017jump}, or efficient method of moments \citep{andersen1999efficient}.} or both returns and option prices/VIX index \citep{kanniainen2014estimating,christoffersen2015option}. Moreover, in academic option pricing research, the standard practice is to evaluate the goodness of models by an option pricing error between market and model prices. This approach implicitly assumes that market prices are {\em always} correct. The inefficiency of option markets, of course, contradicts the use of these metrics. However, we find that options are not priced so incorrectly that one could earn abnormal returns with simple option trading strategies. Thus, we call more research to study efficiency in option markets and possible issues related to the use of option market data for the estimation of option pricing models.


\section*{References}
\bibliography{References}

\begin{thebibliography}{53}
\expandafter\ifx\csname natexlab\endcsname\relax\def\natexlab#1{#1}\fi
\providecommand{\url}[1]{\texttt{#1}}
\providecommand{\href}[2]{#2}
\providecommand{\path}[1]{#1}
\providecommand{\DOIprefix}{doi:}
\providecommand{\ArXivprefix}{arXiv:}
\providecommand{\URLprefix}{URL: }
\providecommand{\Pubmedprefix}{pmid:}
\providecommand{\doi}[1]{\href{http://dx.doi.org/#1}{\path{#1}}}
\providecommand{\Pubmed}[1]{\href{pmid:#1}{\path{#1}}}
\providecommand{\bibinfo}[2]{#2}
\ifx\xfnm\relax \def\xfnm[#1]{\unskip,\space#1}\fi
\bibitem[{Ackert and Tian(2001)}]{ackert2001efficiency}
\bibinfo{author}{Ackert, L.F.}, \bibinfo{author}{Tian, Y.S.},
  \bibinfo{year}{2001}.
\newblock \bibinfo{title}{Efficiency in index options markets and trading in
  stock baskets}.
\newblock \bibinfo{journal}{Journal of banking \& finance}
  \bibinfo{volume}{25}, \bibinfo{pages}{1607--1634}.
\bibitem[{A{\"\i}t-Sahalia and Jacod(2009)}]{ait2009testing}
\bibinfo{author}{A{\"\i}t-Sahalia, Y.}, \bibinfo{author}{Jacod, J.},
  \bibinfo{year}{2009}.
\newblock \bibinfo{title}{Testing for jumps in a discretely observed process}.
\newblock \bibinfo{journal}{The Annals of Statistics} ,
  \bibinfo{pages}{184--222}.
\bibitem[{A{\"\i}t-Sahalia and Kimmel(2007)}]{ai2007maximum}
\bibinfo{author}{A{\"\i}t-Sahalia, Y.}, \bibinfo{author}{Kimmel, R.},
  \bibinfo{year}{2007}.
\newblock \bibinfo{title}{Maximum likelihood estimation of stochastic
  volatility models}.
\newblock \bibinfo{journal}{Journal of financial economics}
  \bibinfo{volume}{83}, \bibinfo{pages}{413--452}.
\bibitem[{A{\"\i}t-Sahalia et~al.(2005)A{\"\i}t-Sahalia, Mykland and
  Zhang}]{ait2005often}
\bibinfo{author}{A{\"\i}t-Sahalia, Y.}, \bibinfo{author}{Mykland, P.A.},
  \bibinfo{author}{Zhang, L.}, \bibinfo{year}{2005}.
\newblock \bibinfo{title}{How often to sample a continuous-time process in the
  presence of market microstructure noise}.
\newblock \bibinfo{journal}{The review of financial studies}
  \bibinfo{volume}{18}, \bibinfo{pages}{351--416}.
\bibitem[{Andersen and Bollerslev(1997)}]{andersen1997intraday}
\bibinfo{author}{Andersen, T.G.}, \bibinfo{author}{Bollerslev, T.},
  \bibinfo{year}{1997}.
\newblock \bibinfo{title}{Intraday periodicity and volatility persistence in
  financial markets}.
\newblock \bibinfo{journal}{Journal of Empirical Finance} \bibinfo{volume}{4},
  \bibinfo{pages}{115--158}.
\bibitem[{Andersen et~al.(2003)Andersen, Bollerslev, Diebold and
  Labys}]{andersen2003modeling}
\bibinfo{author}{Andersen, T.G.}, \bibinfo{author}{Bollerslev, T.},
  \bibinfo{author}{Diebold, F.X.}, \bibinfo{author}{Labys, P.},
  \bibinfo{year}{2003}.
\newblock \bibinfo{title}{Modeling and forecasting realized volatility}.
\newblock \bibinfo{journal}{Econometrica} \bibinfo{volume}{71},
  \bibinfo{pages}{579--625}.
\bibitem[{Andersen et~al.(1999)Andersen, Chung and
  S{\o}rensen}]{andersen1999efficient}
\bibinfo{author}{Andersen, T.G.}, \bibinfo{author}{Chung, H.J.},
  \bibinfo{author}{S{\o}rensen, B.E.}, \bibinfo{year}{1999}.
\newblock \bibinfo{title}{Efficient method of moments estimation of a
  stochastic volatility model: A monte carlo study}.
\newblock \bibinfo{journal}{Journal of econometrics} \bibinfo{volume}{91},
  \bibinfo{pages}{61--87}.
\bibitem[{Balland(2002)}]{balland2002}
\bibinfo{author}{Balland, P.}, \bibinfo{year}{2002}.
\newblock \bibinfo{title}{Deterministic implied volatility models}.
\newblock \bibinfo{journal}{Quantitative Finance} \bibinfo{volume}{2},
  \bibinfo{pages}{31--44}.
\bibitem[{Bandi and Ren{\`o}(2016)}]{bandi2016price}
\bibinfo{author}{Bandi, F.M.}, \bibinfo{author}{Ren{\`o}, R.},
  \bibinfo{year}{2016}.
\newblock \bibinfo{title}{Price and volatility co-jumps}.
\newblock \bibinfo{journal}{Journal of Financial Economics}
  \bibinfo{volume}{119}, \bibinfo{pages}{107--146}.
\bibitem[{Barndorff-Nielsen and Shephard(2004)}]{barndorff2004power}
\bibinfo{author}{Barndorff-Nielsen, O.E.}, \bibinfo{author}{Shephard, N.},
  \bibinfo{year}{2004}.
\newblock \bibinfo{title}{Power and bipower variation with stochastic
  volatility and jumps}.
\newblock \bibinfo{journal}{Journal of Financial Econometrics}
  \bibinfo{volume}{2}, \bibinfo{pages}{1--37}.
\bibitem[{Bernales et~al.(2016)Bernales, Verousis and
  Voukelatos}]{bernales2016investors}
\bibinfo{author}{Bernales, A.}, \bibinfo{author}{Verousis, T.},
  \bibinfo{author}{Voukelatos, N.}, \bibinfo{year}{2016}.
\newblock \bibinfo{title}{Do investors follow the herd in option markets?}
\newblock \bibinfo{journal}{Journal of Banking \& Finance} .
\bibitem[{Binder(1998)}]{binder1998event}
\bibinfo{author}{Binder, J.}, \bibinfo{year}{1998}.
\newblock \bibinfo{title}{The event study methodology since 1969}.
\newblock \bibinfo{journal}{Review of Quantitative Finance and Accounting}
  \bibinfo{volume}{11}, \bibinfo{pages}{111--137}.
\bibitem[{Bollerslev et~al.(2006)Bollerslev, Litvinova and
  Tauchen}]{bollerslev2006leverage}
\bibinfo{author}{Bollerslev, T.}, \bibinfo{author}{Litvinova, J.},
  \bibinfo{author}{Tauchen, G.}, \bibinfo{year}{2006}.
\newblock \bibinfo{title}{Leverage and volatility feedback effects in
  high-frequency data}.
\newblock \bibinfo{journal}{Journal of Financial Econometrics}
  \bibinfo{volume}{4}, \bibinfo{pages}{353--384}.
\bibitem[{Brenner and Subrahmanyam(1988)}]{brenner1988simple}
\bibinfo{author}{Brenner, M.}, \bibinfo{author}{Subrahmanyam, M.G.},
  \bibinfo{year}{1988}.
\newblock \bibinfo{title}{A simple formula to compute the implied standard
  deviation}.
\newblock \bibinfo{journal}{Financial Analysts Journal} ,
  \bibinfo{pages}{80--83}.
\bibitem[{Campbell and Shiller(1987)}]{campbell1987cointegration}
\bibinfo{author}{Campbell, J.Y.}, \bibinfo{author}{Shiller, R.J.},
  \bibinfo{year}{1987}.
\newblock \bibinfo{title}{Cointegration and tests of present value models}.
\newblock \bibinfo{journal}{Journal of political economy} \bibinfo{volume}{95},
  \bibinfo{pages}{1062--1088}.
\bibitem[{Carr and Wu(2006)}]{carr2006tale}
\bibinfo{author}{Carr, P.}, \bibinfo{author}{Wu, L.}, \bibinfo{year}{2006}.
\newblock \bibinfo{title}{A tale of two indices}.
\newblock \bibinfo{journal}{Journal of Derivatives} \bibinfo{volume}{13},
  \bibinfo{pages}{13--29}.
\bibitem[{Cavallo and Mammola(2000)}]{cavallo2000empirical}
\bibinfo{author}{Cavallo, L.}, \bibinfo{author}{Mammola, P.},
  \bibinfo{year}{2000}.
\newblock \bibinfo{title}{Empirical tests of efficiency of the italian index
  options market}.
\newblock \bibinfo{journal}{Journal of Empirical Finance} \bibinfo{volume}{7},
  \bibinfo{pages}{173--193}.
\bibitem[{Christoffersen et~al.(2015)Christoffersen, Feunou and
  Jeon}]{christoffersen2015option}
\bibinfo{author}{Christoffersen, P.}, \bibinfo{author}{Feunou, B.},
  \bibinfo{author}{Jeon, Y.}, \bibinfo{year}{2015}.
\newblock \bibinfo{title}{Option valuation with observable volatility and jump
  dynamics}.
\newblock \bibinfo{journal}{Journal of Banking \& Finance}
  \bibinfo{volume}{61}, \bibinfo{pages}{S101--S120}.
\bibitem[{Chung and Chuwonganant(2018)}]{chung2018market}
\bibinfo{author}{Chung, K.H.}, \bibinfo{author}{Chuwonganant, C.},
  \bibinfo{year}{2018}.
\newblock \bibinfo{title}{Market volatility and stock returns: The role of
  liquidity providers}.
\newblock \bibinfo{journal}{Journal of Financial Markets} \bibinfo{volume}{37},
  \bibinfo{pages}{17--34}.
\bibitem[{Cont and da~Fonseca(2002)}]{cont2002}
\bibinfo{author}{Cont, R.}, \bibinfo{author}{da~Fonseca, J.},
  \bibinfo{year}{2002}.
\newblock \bibinfo{title}{Dynamics of implied volatility surfaces}.
\newblock \bibinfo{journal}{Quantitative Finance} \bibinfo{volume}{2},
  \bibinfo{pages}{45--60}.
\bibitem[{Cont et~al.(2002)Cont, Fonseca and Durrleman}]{cont2002stochastic}
\bibinfo{author}{Cont, R.}, \bibinfo{author}{Fonseca, J.d.},
  \bibinfo{author}{Durrleman, V.}, \bibinfo{year}{2002}.
\newblock \bibinfo{title}{Stochastic models of implied volatility surfaces}.
\newblock \bibinfo{journal}{Economic Notes} \bibinfo{volume}{31},
  \bibinfo{pages}{361--377}.
\bibitem[{De~Bondt and Thaler(1985)}]{de1985does}
\bibinfo{author}{De~Bondt, W.F.}, \bibinfo{author}{Thaler, R.},
  \bibinfo{year}{1985}.
\newblock \bibinfo{title}{Does the stock market overreact?}
\newblock \bibinfo{journal}{The Journal of finance} \bibinfo{volume}{40},
  \bibinfo{pages}{793--805}.
\bibitem[{Deville and Riva(2007)}]{deville2007liquidity}
\bibinfo{author}{Deville, L.}, \bibinfo{author}{Riva, F.},
  \bibinfo{year}{2007}.
\newblock \bibinfo{title}{Liquidity and arbitrage in options markets: A
  survival analysis approach}.
\newblock \bibinfo{journal}{Review of Finance} \bibinfo{volume}{11},
  \bibinfo{pages}{497--525}.
\bibitem[{Ederington and Lee(1993)}]{ederington1993markets}
\bibinfo{author}{Ederington, L.H.}, \bibinfo{author}{Lee, J.H.},
  \bibinfo{year}{1993}.
\newblock \bibinfo{title}{How markets process information: News releases and
  volatility}.
\newblock \bibinfo{journal}{The Journal of Finance} \bibinfo{volume}{48},
  \bibinfo{pages}{1161--1191}.
\bibitem[{Eraker(2004)}]{eraker2004stock}
\bibinfo{author}{Eraker, B.}, \bibinfo{year}{2004}.
\newblock \bibinfo{title}{Do stock prices and volatility jump? reconciling
  evidence from spot and option prices}.
\newblock \bibinfo{journal}{Journal of Finance} \bibinfo{volume}{59},
  \bibinfo{pages}{1367--1403}.
\bibitem[{Eraker et~al.(2003)Eraker, Johannes and Polson}]{eraker2003impact}
\bibinfo{author}{Eraker, B.}, \bibinfo{author}{Johannes, M.},
  \bibinfo{author}{Polson, N.}, \bibinfo{year}{2003}.
\newblock \bibinfo{title}{The impact of jumps in volatility and returns}.
\newblock \bibinfo{journal}{The Journal of Finance} \bibinfo{volume}{58},
  \bibinfo{pages}{1269--1300}.
\bibitem[{Fengler(2006)}]{fengler2006}
\bibinfo{author}{Fengler, M.R.}, \bibinfo{year}{2006}.
\newblock \bibinfo{title}{Semiparametric modeling of implied volatility}.
\newblock \bibinfo{publisher}{Springer Science \& Business Media}.
\bibitem[{Fengler et~al.(2003)Fengler, H{\"a}rdle and Villa}]{fengler2003}
\bibinfo{author}{Fengler, M.R.}, \bibinfo{author}{H{\"a}rdle, W.K.},
  \bibinfo{author}{Villa, C.}, \bibinfo{year}{2003}.
\newblock \bibinfo{title}{The dynamics of implied volatilities: A common
  principal components approach}.
\newblock \bibinfo{journal}{Review of Derivatives Research}
  \bibinfo{volume}{6}, \bibinfo{pages}{179--202}.
\bibitem[{Gatheral(2011)}]{gatheral2011volatility}
\bibinfo{author}{Gatheral, J.}, \bibinfo{year}{2011}.
\newblock \bibinfo{title}{The volatility surface: a practitioner's guide}.
  volume \bibinfo{volume}{357}.
\newblock \bibinfo{publisher}{John Wiley \& Sons}.
\bibitem[{Harvey and Whaley(1992)}]{harvey1992market}
\bibinfo{author}{Harvey, C.R.}, \bibinfo{author}{Whaley, R.E.},
  \bibinfo{year}{1992}.
\newblock \bibinfo{title}{Market volatility prediction and the efficiency of
  the s\&p 100 index option market}.
\newblock \bibinfo{journal}{Journal of Financial Economics}
  \bibinfo{volume}{31}, \bibinfo{pages}{43--73}.
\bibitem[{Heynen(1994)}]{heynen1994empirical}
\bibinfo{author}{Heynen, R.}, \bibinfo{year}{1994}.
\newblock \bibinfo{title}{An empirical investigation of observed smile
  patterns}.
\newblock \bibinfo{journal}{Review of Futures Markets} \bibinfo{volume}{13},
  \bibinfo{pages}{317--317}.
\bibitem[{Izenman(2008)}]{izenman2008}
\bibinfo{author}{Izenman, A.J.}, \bibinfo{year}{2008}.
\newblock \bibinfo{title}{Modern multivariate statistical techniques}.
  volume~\bibinfo{volume}{1}.
\newblock \bibinfo{publisher}{Springer}.
\bibitem[{Jacod et~al.(2010)Jacod, Todorov et~al.}]{jacod2010price}
\bibinfo{author}{Jacod, J.}, \bibinfo{author}{Todorov, V.}, et~al.,
  \bibinfo{year}{2010}.
\newblock \bibinfo{title}{Do price and volatility jump together?}
\newblock \bibinfo{journal}{The Annals of Applied Probability}
  \bibinfo{volume}{20}, \bibinfo{pages}{1425--1469}.
\bibitem[{Jones et~al.(1998)Jones, Lamont and
  Lumsdaine}]{jones1998macroeconomic}
\bibinfo{author}{Jones, C.M.}, \bibinfo{author}{Lamont, O.},
  \bibinfo{author}{Lumsdaine, R.L.}, \bibinfo{year}{1998}.
\newblock \bibinfo{title}{Macroeconomic news and bond market volatility}.
\newblock \bibinfo{journal}{Journal of Financial Economics}
  \bibinfo{volume}{47}, \bibinfo{pages}{315--337}.
\bibitem[{Kanniainen et~al.(2014)Kanniainen, Lin and
  Yang}]{kanniainen2014estimating}
\bibinfo{author}{Kanniainen, J.}, \bibinfo{author}{Lin, B.},
  \bibinfo{author}{Yang, H.}, \bibinfo{year}{2014}.
\newblock \bibinfo{title}{Estimating and using {GARCH} models with {VIX} data
  for option valuation}.
\newblock \bibinfo{journal}{Journal of Banking \& Finance}
  \bibinfo{volume}{43}, \bibinfo{pages}{200--211}.
\bibitem[{Kanniainen and Pich{\'e}(2013)}]{kanniainen2013stock}
\bibinfo{author}{Kanniainen, J.}, \bibinfo{author}{Pich{\'e}, R.},
  \bibinfo{year}{2013}.
\newblock \bibinfo{title}{Stock price dynamics and option valuations under
  volatility feedback effect}.
\newblock \bibinfo{journal}{Physica A: Statistical Mechanics and its
  Applications} \bibinfo{volume}{392}, \bibinfo{pages}{722--740}.
\bibitem[{Kanniainen and Yue(2017)}]{kanniainen2017arrival}
\bibinfo{author}{Kanniainen, J.}, \bibinfo{author}{Yue, Y.},
  \bibinfo{year}{2017}.
\newblock \bibinfo{title}{The arrival of news and jumps in stock markets}.
\newblock \bibinfo{journal}{SSRN Working Paper} .
\bibitem[{Lee(2011)}]{lee2011jumps}
\bibinfo{author}{Lee, S.S.}, \bibinfo{year}{2011}.
\newblock \bibinfo{title}{Jumps and information flow in financial markets}.
\newblock \bibinfo{journal}{Review of Financial Studies} \bibinfo{volume}{25},
  \bibinfo{pages}{439--479}.
\bibitem[{Lee and Mykland(2008)}]{lee2008jumps}
\bibinfo{author}{Lee, S.S.}, \bibinfo{author}{Mykland, P.A.},
  \bibinfo{year}{2008}.
\newblock \bibinfo{title}{Jumps in financial markets: A new nonparametric test
  and jump dynamics}.
\newblock \bibinfo{journal}{Review of Financial Studies} \bibinfo{volume}{21},
  \bibinfo{pages}{2535--2563}.
\bibitem[{McAleer and Medeiros(2008)}]{mcaleer2008realized}
\bibinfo{author}{McAleer, M.}, \bibinfo{author}{Medeiros, M.C.},
  \bibinfo{year}{2008}.
\newblock \bibinfo{title}{Realized volatility: A review}.
\newblock \bibinfo{journal}{Econometric Reviews} \bibinfo{volume}{27},
  \bibinfo{pages}{10--45}.
\bibitem[{Panigirtzoglou and Skiadopoulos(2004)}]{panigirtzoglou2004}
\bibinfo{author}{Panigirtzoglou, N.}, \bibinfo{author}{Skiadopoulos, G.},
  \bibinfo{year}{2004}.
\newblock \bibinfo{title}{A new approach to modeling the dynamics of implied
  distributions: Theory and evidence from the {S}\&{P} 500 options}.
\newblock \bibinfo{journal}{Journal of Banking \& Finance}
  \bibinfo{volume}{28}, \bibinfo{pages}{1499--1520}.
\bibitem[{Poon and Pope(2000)}]{poon2000trading}
\bibinfo{author}{Poon, S.H.}, \bibinfo{author}{Pope, Peter, F.},
  \bibinfo{year}{2000}.
\newblock \bibinfo{title}{Trading volatility spreads: a test of index option
  market efficiency}.
\newblock \bibinfo{journal}{European Financial Management} \bibinfo{volume}{6},
  \bibinfo{pages}{235--260}.
\bibitem[{Rubinstein(1994)}]{rubinstein1994implied}
\bibinfo{author}{Rubinstein, M.}, \bibinfo{year}{1994}.
\newblock \bibinfo{title}{Implied binomial trees}.
\newblock \bibinfo{journal}{The Journal of Finance} \bibinfo{volume}{49},
  \bibinfo{pages}{771--818}.
\bibitem[{Shiller(1981)}]{shiller1981stock}
\bibinfo{author}{Shiller, R.J.}, \bibinfo{year}{1981}.
\newblock \bibinfo{title}{Do stock prices move too much to be justified by
  subsequent changes in dividends?}
\newblock \bibinfo{journal}{The American Economic Review} \bibinfo{volume}{71},
  \bibinfo{pages}{421--436}.
\bibitem[{Siikanen et~al.(2017a)Siikanen, Kanniainen and
  Luoma}]{siikanen2017drives}
\bibinfo{author}{Siikanen, M.}, \bibinfo{author}{Kanniainen, J.},
  \bibinfo{author}{Luoma, A.}, \bibinfo{year}{2017}a.
\newblock \bibinfo{title}{What drives the sensitivity of limit order books to
  company announcement arrivals?}
\newblock \bibinfo{journal}{Economics Letters} \bibinfo{volume}{159},
  \bibinfo{pages}{65--68}.
\bibitem[{Siikanen et~al.(2017b)Siikanen, Kanniainen and
  Valli}]{siikanen2017limit}
\bibinfo{author}{Siikanen, M.}, \bibinfo{author}{Kanniainen, J.},
  \bibinfo{author}{Valli, J.}, \bibinfo{year}{2017}b.
\newblock \bibinfo{title}{Limit order books and liquidity around scheduled and
  non-scheduled announcements: Empirical evidence from nasdaq nordic}.
\newblock \bibinfo{journal}{Finance Research Letters} \bibinfo{volume}{21},
  \bibinfo{pages}{264--271}.
\bibitem[{Skiadopoulos et~al.(2000)Skiadopoulos, Hodges and
  Clewlow}]{skiadopoulos2000}
\bibinfo{author}{Skiadopoulos, G.}, \bibinfo{author}{Hodges, S.},
  \bibinfo{author}{Clewlow, L.}, \bibinfo{year}{2000}.
\newblock \bibinfo{title}{The dynamics of the {S}\&{P} 500 implied volatility
  surface}.
\newblock \bibinfo{journal}{Review of Derivatives Research}
  \bibinfo{volume}{3}, \bibinfo{pages}{263--282}.
\bibitem[{Stephan and Whaley(1990)}]{stephan1990intraday}
\bibinfo{author}{Stephan, J.A.}, \bibinfo{author}{Whaley, R.E.},
  \bibinfo{year}{1990}.
\newblock \bibinfo{title}{Intraday price change and trading volume relations in
  the stock and stock option markets}.
\newblock \bibinfo{journal}{The Journal of Finance} \bibinfo{volume}{45},
  \bibinfo{pages}{191--220}.
\bibitem[{Todorov and Tauchen(2011)}]{todorov2011volatility}
\bibinfo{author}{Todorov, V.}, \bibinfo{author}{Tauchen, G.},
  \bibinfo{year}{2011}.
\newblock \bibinfo{title}{Volatility jumps}.
\newblock \bibinfo{journal}{Journal of Business \& Economic Statistics}
  \bibinfo{volume}{29}, \bibinfo{pages}{356--371}.
\bibitem[{Wahba(1990)}]{wahba1990spline}
\bibinfo{author}{Wahba, G.}, \bibinfo{year}{1990}.
\newblock \bibinfo{title}{Spline models for observational data}.
  volume~\bibinfo{volume}{59}.
\newblock \bibinfo{publisher}{Siam}.
\bibitem[{Xu and Taylor(1994)}]{xu1994magnitude}
\bibinfo{author}{Xu, G.}, \bibinfo{author}{Taylor, S.}, \bibinfo{year}{1994}.
\newblock \bibinfo{title}{The magnitude of implied volatility smiles: Theory
  and empirical evidence for exchange rates}.
\newblock \bibinfo{journal}{Review of Futures Markets} \bibinfo{volume}{13},
  \bibinfo{pages}{355--380}.
\bibitem[{Yang and Kanniainen(2017)}]{yang2017jump}
\bibinfo{author}{Yang, H.}, \bibinfo{author}{Kanniainen, J.},
  \bibinfo{year}{2017}.
\newblock \bibinfo{title}{Jump and volatility dynamics for the {S\&P} 500:
  Evidence for infinite-activity jumps with non-affine volatility dynamics from
  stock and option markets}.
\newblock \bibinfo{journal}{Review of Finance} \bibinfo{volume}{21},
  \bibinfo{pages}{811--844}.
\bibitem[{Zhang et~al.(2005)Zhang, Mykland and
  A{\"\i}t-Sahalia}]{zhang2005tale}
\bibinfo{author}{Zhang, L.}, \bibinfo{author}{Mykland, P.A.},
  \bibinfo{author}{A{\"\i}t-Sahalia, Y.}, \bibinfo{year}{2005}.
\newblock \bibinfo{title}{A tale of two time scales: Determining integrated
  volatility with noisy high-frequency data}.
\newblock \bibinfo{journal}{Journal of the American Statistical Association}
  \bibinfo{volume}{100}, \bibinfo{pages}{1394--1411}.

\end{thebibliography}

\newpage

\section*{Appendix}
\begin{figure}[b!]
\centering
\includegraphics[scale=0.65]{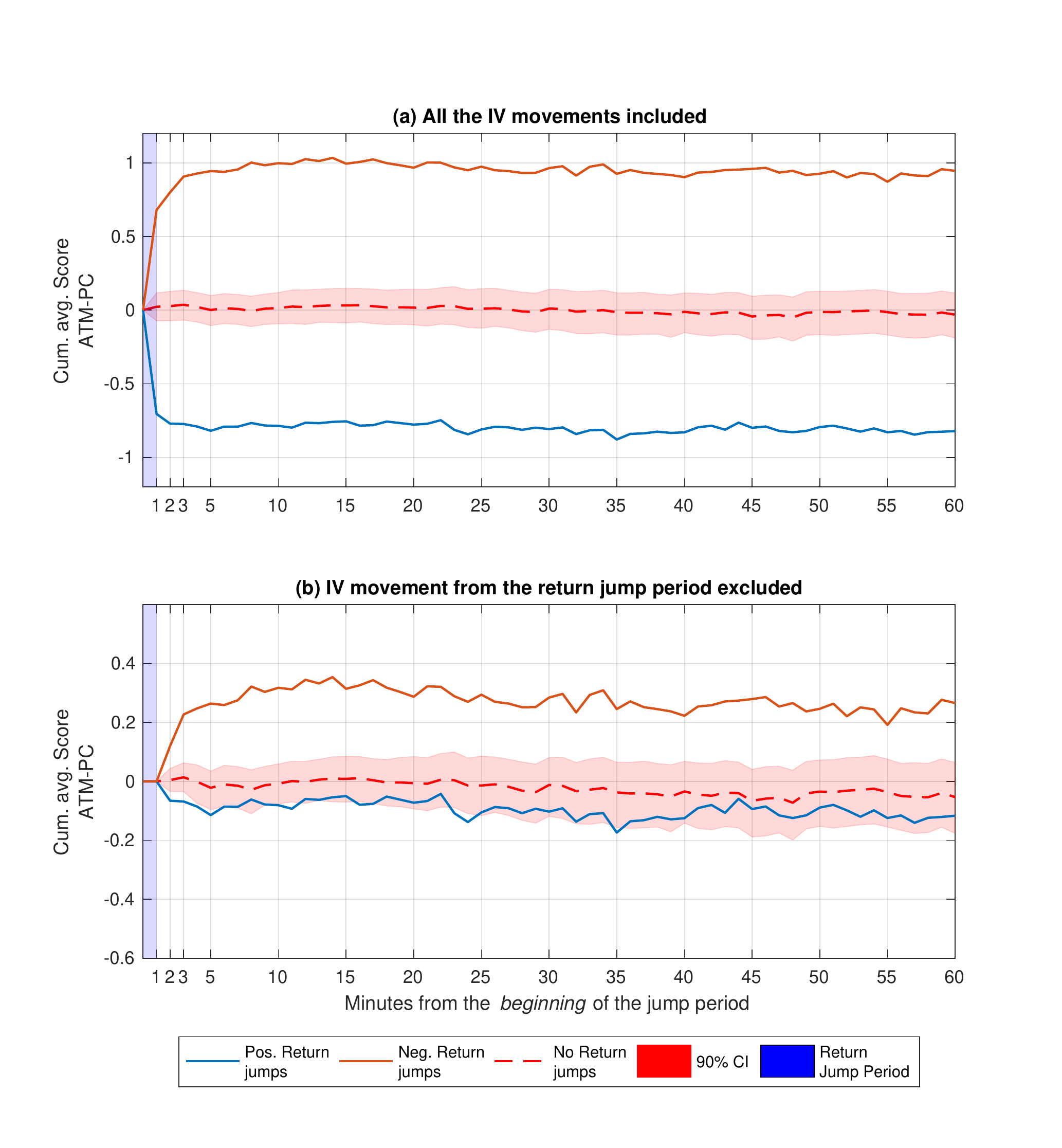}
\caption{Average cumulative movements in implied volatility (IV) after the arrival of return jumps. Here the IV is measured by {\em at-the-money principal component}. Time index refers to minutes from the beginning of the 1-minute return jump period. That is, the return jump period is between 0 and 1 minutes (blue interval), during which jumps have arrived, and the post-movements in the IV are from minute 1 onwards. Cumulative IV movements after positive jumps are in blue and after negative jumps in red. The dashed curve represents cumulative IV movements when no return jumps have arrived. Panel (a) plots cumulative IV, including the IV movement, over the return jump period. Panel (b) excludes the simultaneous IV movement with the return jump. The 90\% confidence interval is obtained by numerical bootstrapping.} 
\label{FIG:IV-Dynamics_ATM_PC_90}
\end{figure}

\begin{figure}
\centering
\includegraphics[scale=0.65]{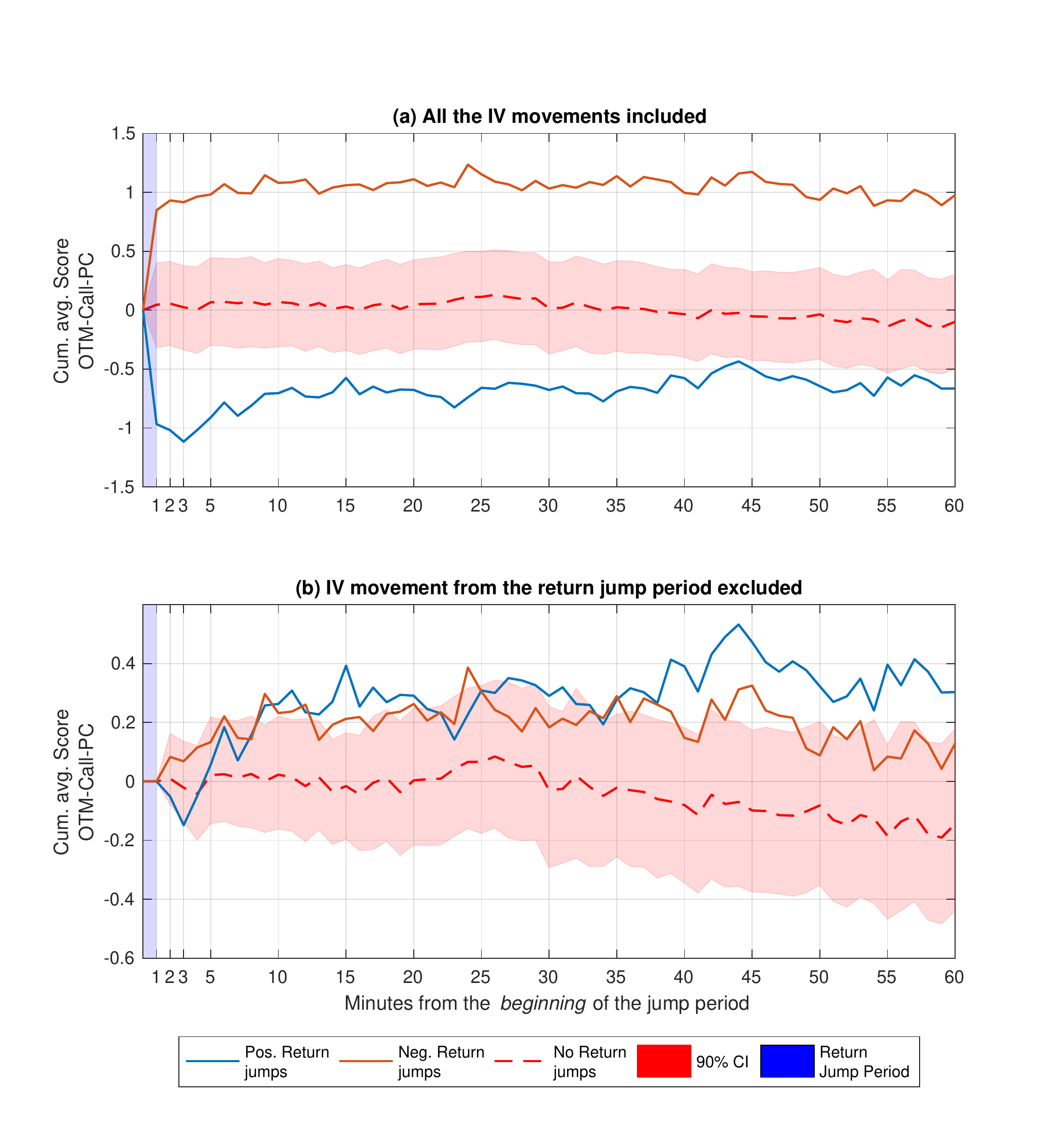}
\caption{Average cumulative movements in implied volatility (IV) after the arrival of return jumps. Here, the IV is measured by {\em out-of-the-money call option principal component}. Time index refers to minutes from the beginning of the 1-minute return jump period. That is, the return jump period is between 0 and 1 minutes (blue interval), during which jumps have arrived, and the post-movements in the IV are from minute 1 onwards. Cumulative IV movements after positive jumps are in blue and after negative jumps in red. The dashed curve represents cumulative IV movements when no return jumps have arrived. Panel (a) plots cumulative IV, including the IV movement, over the return jump period. Panel (b) excludes the simultaneous IV movement with the return jump. The 90\% confidence interval is obtained by numerical bootstrapping.} 
\label{FIG:IV-Dynamics_OTM_Call_PC_90}
\end{figure}

\begin{figure}
\centering
\includegraphics[scale=0.65]{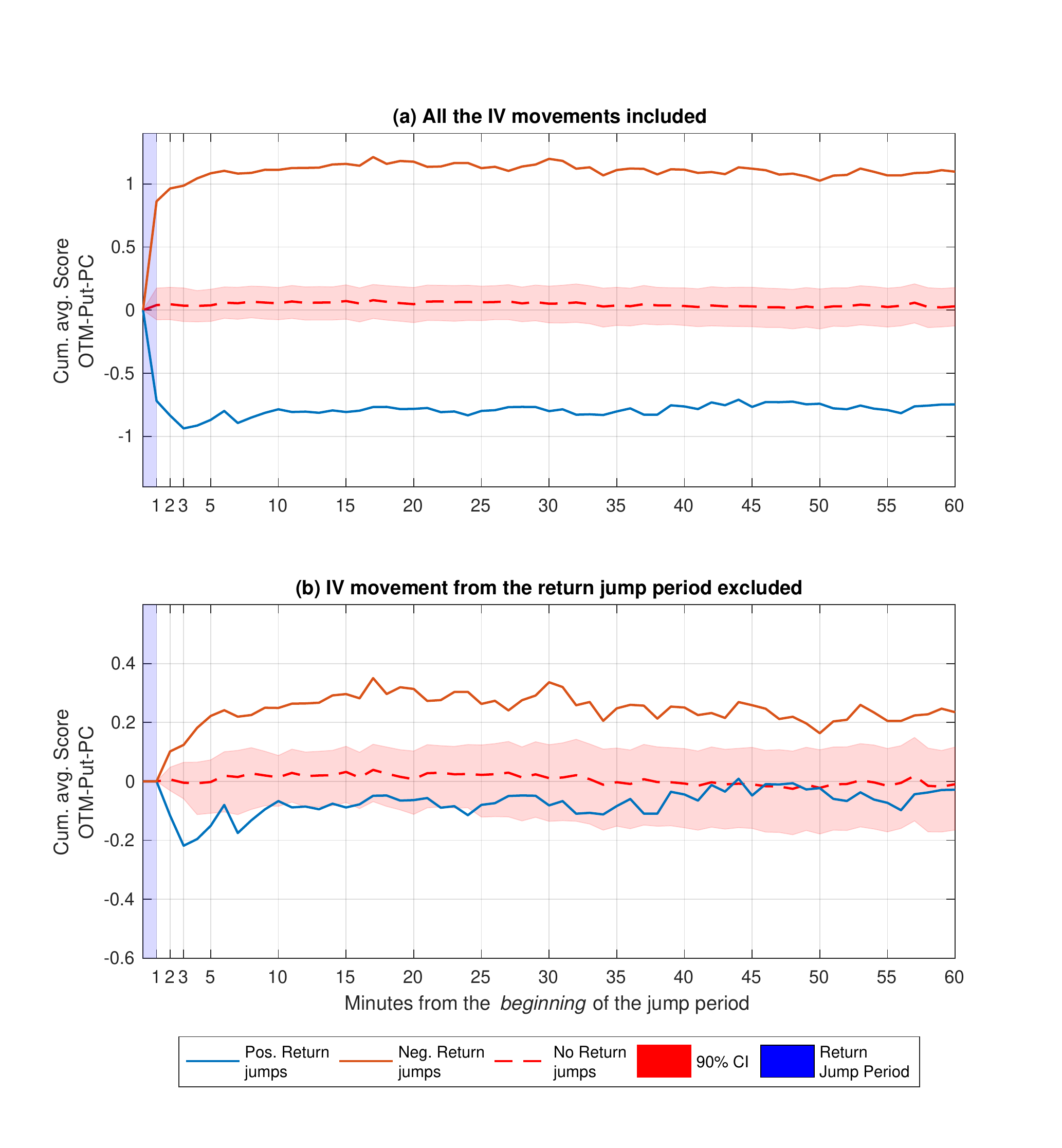}
\caption{Average cumulative movements in implied volatility (IV) after the arrival of return jumps. Here, the IV is measured by {\em out-of-the-money put option principal component}. Time index refers to minutes from the beginning of the 1-minute return jump period. That is, the return jump period is between 0 and 1 minutes (blue interval), during which jumps have arrived, and the post-movements in the IV are from minute 1 onwards. Cumulative IV movements after positive jumps are in blue, and after negative jumps are in red. The dashed curve represents cumulative IV movements when no return jumps have arrived. Panel (a) plots cumulative IV, including the IV movement, over the return jump period. Panel (b) excludes the simultaneous IV movement with the return jump. The 90\% confidence interval is obtained by numerical bootstrapping.} 
\label{FIG:IV-Dynamics_OTM-Put_PC_90}
\end{figure}

\begin{sidewaystable}
\caption{Regression results {\em excluding} the IV moments observed from the return jump window. In this table, the parameter estimates for $\beta_p$ and $\beta_n$ are reported for various IV window lengths from 5 to 60 minutes. Data is observed from 09:30 to 10:30, and return jumps are detected using 1-minute windows with a significance level of 1\%. The complete regression results are available upon request.}\label{TAB:IV_Regression_Results_Exc_First_Reaction_5_60}
\centering
\resizebox{1\textheight}{!}
{
\begin{tabular}{lccccc|ccccc}
      & \multicolumn{5}{c|}{\boldmath{}\textbf{$\beta_p$}\unboldmath{}} & \multicolumn{5}{c}{\boldmath{}\textbf{$\beta_n$}\unboldmath{}} \\
\textbf{Minutes} & \textbf{5} & \textbf{15} & \textbf{20} & \textbf{30} & \textbf{60} & \textbf{5} & \textbf{15} & 
\textbf{20} & \textbf{30} & \textbf{60} \\
\hline
\textbf{ATM-IV} &       &       &       &       &       &       &       &       &       &  \\
\textbf{3 months} & -0.096 & -0.085 & -0.079 & -0.070 & -0.019 & 0.136 & 0.159 & 0.163 & 0.191 & 0.218 \\
      & (1.123E-03**) & (0.019*) & (0.042*) & (0.122) & (0.720) & (9.822E-06***) & (2.780E-05***) & (5.683E-05***) & (5.258E-05***) & (1.055E-04***) \\
\textbf{6 months} & -0.055 & -0.050 & -0.047 & -0.064 & -7.963E-03 & 0.080 & 0.101 & 0.113 & 0.121 & 0.149 \\
      & (1.770E-03**) & (0.012*) & (0.030*) & (0.014*) & (0.820) & (1.168E-05***) & (1.140E-06***) & (7.884E-07***) & (6.987E-06***) & (4.081E-05***) \\
\textbf{9 months} & -0.044 & -0.021 & -0.037 & -0.046 & -0.028 & 0.062 & 0.099 & 0.091 & 0.113 & 0.109 \\
      & (1.834E-04***) & (0.199) & (0.033*) & (0.025*) & (0.308) & (3.843E-07***) & (1.250E-08***) & (5.649E-07***) & (1.223E-07***) & (1.319E-04***) \\
\textbf{ATM-PC} &       &       &       &       &       &       &       &       &       &  \\
\textbf{3 months} & -0.092 & -0.057 & -0.104 & -0.156 & -0.048 & 0.270 & 0.317 & 0.290 & 0.257 & 0.367 \\
      & (0.050*) & (0.314) & (0.110) & (0.037*) & (0.618) & (3.722E-08***) & (1.062E-07***) & (2.111E-05***) & (9.349E-04***) & (2.634E-04***) \\
\textbf{6 months} & -0.115 & -0.081 & -0.045 & -0.122 & -0.049 & 0.160 & 0.210 & 0.253 & 0.188 & 0.227 \\
      & (3.862E-05***) & (0.024*) & (0.273) & (0.010*) & (0.451) & (3.489E-08***) & (1.961E-08***) & (5.572E-09***) & (1.431E-04***) & (7.074E-04***) \\
\textbf{9 months} & -0.097 & -0.083 & -0.072 & -0.063 & -0.073 & 0.104 & 0.146 & 0.162 & 0.235 & 0.186 \\
      & (1.474E-05***) & (9.979E-03**) & (0.036*) & (0.114) & (0.181) & (7.651E-06***) & (1.518E-05***) & (6.975E-06***) & (2.086E-08***) & (1.008E-03**) \\
\textbf{OTM-Call-PC} &       &       &       &       &       &       &       &       &       &  \\
\textbf{3 months} & 0.061 & 0.451 & 0.346 & 0.240 & 0.459 & 0.171 & 0.376 & 0.390 & 0.175 & 0.338 \\
      & (0.630) & (2.770E-03**) & (0.040*) & (0.200) & (0.041*) & (0.196) & (0.016*) & (0.026*) & (0.369) & (0.146) \\
\textbf{6 months} & -0.058 & 0.124 & 0.118 & 0.080 & 0.161 & 0.055 & 0.143 & 0.145 & 0.178 & 0.218 \\
      & (0.311) & (0.097*) & (0.136) & (0.359) & (0.139) & (0.348) & (0.066*) & (0.076*) & (0.050*) & (0.053*) \\
\textbf{9 months} & -0.017 & 0.026 & 0.031 & 7.318E-03 & 0.073 & 0.094 & 0.135 & 0.130 & 0.159 & 0.149 \\
      & (0.679) & (0.581) & (0.550) & (0.900) & (0.290) & (0.024*) & (5.093E-03**) & (0.017*) & (8.726E-03**) & (0.039*) \\
\textbf{OTM-Put-PC} &       &       &       &       &       &       &       &       &       &  \\
\textbf{3 months} & -0.148 & -0.122 & -0.098 & -0.066 & 0.026 & 0.176 & 0.241 & 0.252 & 0.309 & 0.248 \\
      & (7.330E-03**) & (0.073*) & (0.186) & (0.450) & (0.794) & (2.163E-03**) & (6.402E-04***) & (1.132E-03**) & (6.503E-04***) & (0.018*) \\
\textbf{6 months} & -0.127 & -0.121 & -0.080 & -0.161 & -0.068 & 0.108 & 0.139 & 0.172 & 0.177 & 0.180 \\
      & (1.252E-04***) & (3.624E-03**) & (0.056*) & (1.683E-03**) & (0.311) & (1.579E-03**) & (1.327E-03**) & (8.416E-05***) & (9.222E-04***) & (9.897E-03**) \\
\textbf{9 months} & -0.055 & -0.068 & -0.065 & -0.093 & -0.017 & 0.145 & 0.117 & 0.151 & 0.146 & 0.247 \\
      & (5.787E-03**) & (0.019*) & (0.030*) & (0.010*) & (0.739) & (5.708E-12***) & (1.045E-04***) & (1.639E-06***) & (1.134E-04***) & (4.223E-06***) \\
\end{tabular}%
}
\end{sidewaystable}

\begin{sidewaystable}
\caption{Robustness check: Jump detection with a significance level of 5\% (instead of 1\%). The regressions {\em exclude} the IV moments observed from the return jump window. Data is observed from 09:30 to 10:30, and return jumps are detected using 1-minute windows. The parameter estimates for $\beta_p$ and $\beta_n$ are reported for various IV window lengths from 5 to 60 minutes. The complete regression results are available upon request.}\label{TAB:IV_Regression_Results_Exc_First_Reaction_5_60_5Pros}
\centering
\resizebox{1\textheight}{!}
{
\begin{tabular}{lccccc|ccccc}
      & \multicolumn{5}{c|}{\boldmath{}\textbf{$\beta_p$}\unboldmath{}} & \multicolumn{5}{c}{\boldmath{}\textbf{$\beta_n$}\unboldmath{}} \\
\textbf{Minutes} & \textbf{5} & \textbf{15} & \textbf{20} & \textbf{30} & \textbf{60} & \textbf{5} & \textbf{15} & \textbf{20} & \textbf{30} & \textbf{60} \\
\hline
\textbf{ATM-IV} &       &       &       &       &       &       &       &       &       &  \\
\textbf{3 months} & -0.094 & -0.132 & -0.070 & -0.080 & -0.032 & 0.125 & 0.106 & 0.152 & 0.156 & 0.187 \\
      & (9.894E-04***) & (2.816E-04***) & (0.066*) & (0.075*) & (0.547) & (2.828E-05***) & (5.526E-03**) & (1.502E-04***) & (9.570E-04***) & (7.739E-04***) \\
\textbf{6 months} & -0.049 & -0.055 & -0.056 & -0.043 & -2.267E-04 & 0.084 & 0.095 & 0.100 & 0.126 & 0.142 \\
      & (3.446E-03**) & (6.320E-03**) & (9.293E-03**) & (0.074*) & (0.994) & (1.564E-06***) & (6.473E-06***) & (8.126E-06***) & (6.890E-07***) & (3.046E-05***) \\
\textbf{9 months} & -0.054 & -0.047 & -0.040 & -0.061 & -0.020 & 0.053 & 0.072 & 0.088 & 0.090 & 0.112 \\
      & (1.025E-06***) & (2.687E-03**) & (0.023*) & (1.341E-03**) & (0.441) & (4.635E-06***) & (8.851E-06***) & (2.363E-06***) & (6.742E-06***) & (2.965E-05***) \\
\textbf{ATM-PC} &       &       &       &       &       &       &       &       &       &  \\
\textbf{3 months} & -0.125 & -6.787E-03 & -0.047 & -0.064 & -0.041 & 0.233 & 0.369 & 0.364 & 0.351 & 0.353 \\
      & (0.012*) & (0.906) & (0.453) & (0.368) & (0.644) & (8.238E-06***) & (9.621E-10***) & (3.606E-08***) & (3.005E-06***) & (1.482E-04***) \\
\textbf{6 months} & -0.092 & -0.062 & -0.065 & -0.100 & -0.082 & 0.196 & 0.238 & 0.239 & 0.212 & 0.187 \\
      & (6.447E-04***) & (0.080*) & (0.094*) & (0.030*) & (0.196) & (4.388E-12***) & (2.008E-10***) & (4.650E-09***) & (1.207E-05***) & (5.038E-03**) \\
\textbf{9 months} & -0.103 & -0.083 & -0.059 & -0.077 & -0.081 & 0.085 & 0.137 & 0.167 & 0.199 & 0.162 \\
      & (1.143E-06***) & (6.342E-03**) & (0.076*) & (0.050*) & (0.125) & (1.230E-04***) & (1.560E-05***) & (1.884E-06***) & (1.629E-06***) & (3.262E-03**) \\
\textbf{OTM-Call-PC} &       &       &       &       &       &       &       &       &       &  \\
\textbf{3 months} & 0.107 & 0.241 & 0.360 & 0.554 & 0.355 & 0.254 & 0.207 & 0.349 & 0.484 & 0.124 \\
      & (0.367) & (0.098*) & (0.038*) & (1.547E-03**) & (0.103) & (0.041*) & (0.175) & (0.054*) & (8.137E-03**) & (0.585) \\
\textbf{6 months} & -0.024 & -0.021 & 0.102 & 0.121 & 0.136 & 0.111 & 0.034 & 0.132 & 0.233 & 0.201 \\
      & (0.681) & (0.759) & (0.181) & (0.138) & (0.188) & (0.064*) & (0.642) & (0.100*) & (6.347E-03**) & (0.062*) \\
\textbf{9 months} & -0.026 & -0.023 & 0.037 & 0.040 & 0.085 & 0.071 & 0.093 & 0.134 & 0.185 & 0.136 \\
      & (0.470) & (0.602) & (0.485) & (0.497) & (0.196) & (0.062*) & (0.047*) & (0.015*) & (2.514E-03**) & (0.049*) \\
\textbf{OTM-Put-PC} &       &       &       &       &       &       &       &       &       &  \\
\textbf{3 months} & -0.117 & -0.162 & -0.052 & -0.068 & -0.022 & 0.198 & 0.208 & 0.280 & 0.302 & 0.196 \\
      & (0.026*) & (0.019*) & (0.491) & (0.425) & (0.833) & (3.381E-04***) & (3.887E-03**) & (4.476E-04***) & (6.715E-04***) & (0.075*) \\
\textbf{6 months} & -0.102 & -0.112 & -0.065 & -0.059 & -0.036 & 0.127 & 0.143 & 0.182 & 0.253 & 0.196 \\
      & (1.617E-03**) & (4.905E-03**) & (0.139) & (0.236) & (0.582) & (1.970E-04***) & (5.813E-04***) & (7.286E-05***) & (1.420E-06***) & (3.828E-03**) \\
\textbf{9 months} & -0.045 & -0.051 & -0.040 & -0.087 & -4.712E-03 & 0.142 & 0.150 & 0.195 & 0.167 & 0.254 \\
      & (0.024*) & (0.073*) & (0.187) & (0.016*) & (0.922) & (2.113E-11***) & (4.662E-07***) & (8.452E-10***) & (1.171E-05***) & (5.809E-07***) \\
\end{tabular}%
}
\end{sidewaystable}

\begin{sidewaystable}
\caption{Robustness check: Observations extracted from 09:30 to 12:30 (instead from 09:30 to 10:30). The regressions {\em exclude} the IV moments observed from the return jump window. Return jumps are detected using 1-minute windows with a significance level of 1\%. The parameter estimates for $\beta_p$ and $\beta_n$ are reported for various IV window lengths from 5 to 60 minutes. The complete regression results are available upon request.}\label{TAB:IV_Regression_Results_Exc_First_Reaction_5_60_1230}
\centering
\resizebox{1\textheight}{!}
{
\begin{tabular}{lccccc|ccccc}
      & \multicolumn{5}{c|}{\boldmath{}\textbf{$\beta_p$}\unboldmath{}} & \multicolumn{5}{c}{\boldmath{}\textbf{$\beta_n$}\unboldmath{}} \\
\textbf{Minutes} & \textbf{5} & \textbf{15} & \textbf{20} & \textbf{30} & \textbf{60} & \textbf{5} & \textbf{15} & \textbf{20} & \textbf{30} & \textbf{60} \\
\hline
\textbf{ATM-IV} &       &       &       &       &       &       &       &       &       &  \\
\textbf{3 months} & -0.068 & -0.097 & -0.099 & -0.087 & -0.038 & 0.164 & 0.151 & 0.145 & 0.176 & 0.186 \\
      & (0.023*) & (7.464E-03**) & (7.915E-03**) & (0.071*) & (0.488) & (1.431E-07***) & (5.666E-05***) & (1.948E-04***) & (4.582E-04***) & (9.708E-04***) \\
\textbf{6 months} & -0.060 & -0.047 & -0.061 & -0.083 & -0.031 & 0.077 & 0.110 & 0.105 & 0.101 & 0.123 \\
      & (1.673E-04***) & (0.023*) & (5.059E-03**) & (1.646E-03**) & (0.345) & (3.229E-06***) & (3.772E-07***) & (4.143E-06***) & (2.210E-04***) & (3.549E-04***) \\
\textbf{9 months} & -0.029 & -0.022 & -0.022 & -0.047 & -0.014 & 0.076 & 0.096 & 0.107 & 0.108 & 0.120 \\
      & (0.016*) & (0.172) & (0.226) & (0.022*) & (0.601) & (1.957E-09***) & (1.170E-08***) & (9.302E-09***) & (4.041E-07***) & (1.075E-05***) \\
\textbf{ATM-PC} &       &       &       &       &       &       &       &       &       &  \\
\textbf{3 months} & -0.100 & -0.057 & -0.107 & -0.165 & -0.132 & 0.257 & 0.309 & 0.276 & 0.242 & 0.268 \\
      & (0.031*) & (0.306) & (0.111) & (0.029*) & (0.150) & (1.165E-07***) & (1.248E-07***) & (7.550E-05***) & (2.101E-03**) & (5.072E-03**) \\
\textbf{6 months} & -0.114 & -0.078 & -0.066 & -0.124 & -0.072 & 0.164 & 0.215 & 0.238 & 0.185 & 0.202 \\
      & (2.487E-05***) & (0.034*) & (0.117) & (0.011*) & (0.279) & (6.795E-09***) & (2.790E-08***) & (6.287E-08***) & (2.852E-04***) & (3.672E-03**) \\
\textbf{9 months} & -0.079 & -0.039 & -0.059 & -0.090 & -0.106 & 0.116 & 0.188 & 0.170 & 0.192 & 0.137 \\
      & (2.397E-04***) & (0.223) & (0.095*) & (0.033*) & (0.045*) & (2.153E-07***) & (2.674E-08***) & (4.154E-06***) & (1.281E-05***) & (0.013*) \\
\textbf{OTM-Call-PC} &       &       &       &       &       &       &       &       &       &  \\
\textbf{3 months} & 0.062 & 0.409 & 0.431 & 0.559 & 0.293 & 0.175 & 0.295 & 0.439 & 0.430 & 0.126 \\
      & (0.593) & (2.822E-03**) & (7.937E-03**) & (1.371E-03**) & (0.173) & (0.146) & (0.038*) & (9.293E-03**) & (0.018*) & (0.572) \\
\textbf{6 months} & -3.830E-03 & 0.115 & 0.111 & 0.131 & 0.258 & 0.118 & 0.112 & 0.112 & 0.192 & 0.286 \\
      & (0.940) & (0.120) & (0.140) & (0.119) & (0.014*) & (0.025*) & (0.149) & (0.152) & (0.028*) & (8.742E-03**) \\
\textbf{9 months} & -0.017 & 0.020 & 0.066 & -8.182E-03 & 0.078 & 0.103 & 0.115 & 0.147 & 0.115 & 0.116 \\
      & (0.633) & (0.674) & (0.189) & (0.888) & (0.235) & (5.188E-03**) & (0.019*) & (4.817E-03**) & (0.059*) & (0.089*) \\
\textbf{OTM-Put-PC} &       &       &       &       &       &       &       &       &       &  \\
\textbf{3 months} & -0.088 & -0.150 & -0.154 & -0.123 & 4.761E-03 & 0.232 & 0.228 & 0.203 & 0.257 & 0.220 \\
      & (0.132) & (0.026*) & (0.029*) & (0.166) & (0.961) & (1.323E-04***) & (1.228E-03**) & (5.788E-03**) & (5.511E-03**) & (0.028*) \\
\textbf{6 months} & -0.142 & -0.155 & -0.144 & -0.153 & -0.064 & 0.091 & 0.125 & 0.127 & 0.187 & 0.188 \\
      & (3.106E-06***) & (1.136E-04***) & (7.460E-04***) & (2.414E-03**) & (0.327) & (3.959E-03**) & (2.739E-03**) & (4.160E-03**) & (3.572E-04***) & (5.964E-03**) \\
\textbf{9 months} & -0.056 & -0.047 & -0.079 & -0.077 & -0.048 & 0.143 & 0.139 & 0.136 & 0.169 & 0.215 \\
      & (5.184E-03**) & (0.108) & (9.344E-03**) & (0.036*) & (0.351) & (1.210E-11***) & (5.562E-06***) & (1.834E-05***) & (8.999E-06***) & (5.499E-05***) \\
\end{tabular}%
}
\end{sidewaystable}

\begin{sidewaystable}
\caption{Robustness check: Regressions results iterated 1,000 times with different realizations of reference data. Different realizations of the reference data result from the procedure where the starting points are drawn from the empirical distribution of return jump locations. Mean, standard deviation, and 2.5\% and 97.5\% quantiles (Q2.5\% and Q97.5\%) are reported for the parameter estimates and p-values. The regressions {\em exclude} the IV moments observed from the return jump window. Data is observed from 09:30 to 10:30, and return jumps are detected using 1-minute windows with a significance level of 1\%. Panel (a) reports results for a 5-minute window and Panel (b) for a 60-minute IV window.}\label{TAB:IV_Regression_Results_Exc_First_Reaction_5_60_1000Reg}
\centering
\resizebox{1\textheight}{!}
{
\begin{tabular}{ll|cccc|cccc|cccc}
\multicolumn{14}{c}{\textbf{Panel (a): 5 minute IV window}} \\
      &       & \multicolumn{4}{c|}{\textbf{3 months}} & \multicolumn{4}{c|}{\textbf{6 months}} & \multicolumn{4}{c}{\textbf{9 months}} \\
      &       & \textbf{Mean} & \textbf{StDev} & \textbf{Q2.5\%} & \textbf{Q97.5\%} & \textbf{Mean} & \textbf{StDev} & \textbf{Q2.5\%} & \textbf{Q97.5\%} & \textbf{Mean} & \textbf{StDev} & \textbf{Q2.5\%} & \textbf{Q97.5\%} \\
\hline
\textbf{ATM-IV} &       &       &       &       &       &       &       &       &       &       &       &       &  \\
$\beta_P$ & Estim. & -0.081 & 0.011 & -0.100 & -0.081 & -0.063 & 5.554E-03 & -0.073 & -0.063 & -0.041 & 4.216E-03 & -0.049 & -0.041 \\
      & P-value & 0.018* & 0.026 & 1.052E-03** & 0.018* & 5.063E-04*** & 1.362E-03 & 6.161E-06*** & 5.063E-04*** & 1.418E-03** & 2.338E-03 & 4.848E-05*** & 1.418E-03** \\
$\beta_N$ & Estim. & 0.151 & 0.011 & 0.133 & 0.151 & 0.073 & 5.674E-03 & 0.062 & 0.073 & 0.065 & 4.479E-03 & 0.057 & 0.065 \\
      & P-value & 8.482E-06*** & 1.430E-05 & 2.235E-07*** & 8.482E-06*** & 5.833E-05*** & 1.343E-04 & 2.085E-06*** & 5.833E-05*** & 1.272E-06*** & 3.724E-06 & 7.151E-09*** & 1.272E-06*** \\
\textbf{ATM-IV-PC} &       &       &       &       &       &       &       &       &       &       &       &       &  \\
$\beta_P$ & Estim. & -0.121 & 0.015 & -0.150 & -0.121 & -0.106 & 8.841E-03 & -0.123 & -0.106 & -0.095 & 7.953E-03 & -0.111 & -0.095 \\
      & P-value & 0.016* & 0.018 & 1.542E-03** & 0.016* & 2.575E-04*** & 4.949E-04 & 6.147E-06*** & 2.575E-04*** & 6.742E-05*** & 1.549E-04 & 1.287E-06*** & 6.742E-05*** \\
$\beta_N$ & Estim. & 0.239 & 0.015 & 0.209 & 0.239 & 0.170 & 9.231E-03 & 0.152 & 0.170 & 0.106 & 8.424E-03 & 0.090 & 0.106 \\
      & P-value & 4.347E-06*** & 8.979E-06 & 1.067E-07*** & 4.347E-06*** & 1.298E-08*** & 6.022E-08 & 5.939E-11*** & 1.298E-08*** & 2.841E-05*** & 9.424E-05 & 1.600E-07*** & 2.841E-05*** \\
\textbf{OTM-Call} &       &       &       &       &       &       &       &       &       &       &       &       &  \\
$\beta_P$ & Estim. & 0.081 & 0.045 & -0.015 & 0.081 & -0.047 & 0.020 & -0.086 & -0.047 & -0.039 & 0.013 & -0.064 & -0.039 \\
      & P-value & 0.503 & 0.216 & 0.153 & 0.503 & 0.433 & 0.191 & 0.146 & 0.433 & 0.320 & 0.152 & 0.101 & 0.320 \\
$\beta_N$ & Estim. & 0.189 & 0.045 & 0.092 & 0.189 & 0.066 & 0.020 & 0.027 & 0.066 & 0.072 & 0.013 & 0.048 & 0.072 \\
      & P-value & 0.147 & 0.116 & 0.022* & 0.147 & 0.297 & 0.154 & 0.074* & 0.297 & 0.081* & 0.060 & 0.013* & 0.081* \\
\textbf{OTM-Put} &       &       &       &       &       &       &       &       &       &       &       &       &  \\
$\beta_P$ & Estim. & -0.126 & 0.022 & -0.168 & -0.126 & -0.124 & 0.011 & -0.144 & -0.124 & -0.057 & 6.967E-03 & -0.070 & -0.057 \\
      & P-value & 0.045* & 0.057 & 2.939E-03** & 0.045* & 3.308E-04*** & 9.697E-04 & 2.913E-06*** & 3.308E-04*** & 9.173E-03** & 0.012 & 5.221E-04*** & 9.173E-03** \\
$\beta_N$ & Estim. & 0.196 & 0.022 & 0.152 & 0.196 & 0.112 & 0.011 & 0.091 & 0.112 & 0.144 & 7.324E-03 & 0.130 & 0.144 \\
      & P-value & 2.267E-03** & 3.684E-03 & 1.301E-04*** & 2.267E-03** & 1.102E-03** & 1.496E-03 & 6.085E-05*** & 1.102E-03** & 1.712E-10*** & 6.461E-10 & 4.075E-13*** & 1.712E-10*** \\
      & \multicolumn{1}{r}{} &       &       &       & \multicolumn{1}{r}{} &       &       &       & \multicolumn{1}{r}{} &       &       &       &  \\
\multicolumn{14}{c}{\textbf{Panel (b): 60 minute IV window}} \\
      &       & \multicolumn{4}{c|}{\textbf{3 months}} & \multicolumn{4}{c|}{\textbf{6 months}} & \multicolumn{4}{c}{\textbf{9 months}} \\
      &       & \textbf{Mean} & \textbf{StDev} & \textbf{Q2.5\%} & \textbf{Q97.5\%} & \textbf{Mean} & \textbf{StDev} & \textbf{Q2.5\%} & \textbf{Q97.5\%} & \textbf{Mean} & \textbf{StDev} & \textbf{Q2.5\%} & \textbf{Q97.5\%} \\
\hline
\textbf{ATM-IV} &       &       &       &       &       &       &       &       &       &       &       &       &  \\
$\beta_P$ & Estim. & -0.030 & 0.018 & -0.064 & -0.030 & -0.025 & 0.012 & -0.049 & -0.025 & -0.027 & 9.890E-03 & -0.046 & -0.027 \\
      & P-value & 0.593 & 0.196 & 0.235 & 0.593 & 0.492 & 0.200 & 0.155 & 0.492 & 0.364 & 0.183 & 0.099* & 0.364 \\
$\beta_N$ & Estim. & 0.202 & 0.019 & 0.165 & 0.202 & 0.131 & 0.013 & 0.105 & 0.131 & 0.109 & 0.011 & 0.089 & 0.109 \\
      & P-value & 7.333E-04*** & 9.645E-04 & 4.297E-05*** & 7.333E-04*** & 5.425E-04*** & 9.895E-04 & 1.458E-05*** & 5.425E-04*** & 3.727E-04*** & 7.086E-04 & 7.390E-06*** & 3.727E-04*** \\
\textbf{ATM-IV-PC} &       &       &       &       &       &       &       &       &       &       &       &       &  \\
$\beta_P$ & Estim. & -0.064 & 0.033 & -0.128 & -0.064 & -0.027 & 0.024 & -0.073 & -0.027 & -0.067 & 0.020 & -0.107 & -0.067 \\
      & P-value & 0.519 & 0.204 & 0.180 & 0.519 & 0.665 & 0.200 & 0.275 & 0.665 & 0.261 & 0.153 & 0.057* & 0.261 \\
$\beta_N$ & Estim. & 0.352 & 0.036 & 0.283 & 0.352 & 0.250 & 0.026 & 0.199 & 0.250 & 0.194 & 0.021 & 0.152 & 0.194 \\
      & P-value & 8.444E-04*** & 1.727E-03 & 2.526E-05*** & 8.444E-04*** & 7.476E-04*** & 1.354E-03 & 1.595E-05*** & 7.476E-04*** & 1.638E-03** & 2.501E-03 & 4.555E-05*** & 1.638E-03** \\
\textbf{OTM-Call} &       &       &       &       &       &       &       &       &       &       &       &       &  \\
$\beta_P$ & Estim. & 0.368 & 0.075 & 0.223 & 0.368 & 0.123 & 0.035 & 0.057 & 0.123 & 0.060 & 0.023 & 0.015 & 0.060 \\
      & P-value & 0.109 & 0.076 & 0.017* & 0.109 & 0.268 & 0.140 & 0.071* & 0.268 & 0.414 & 0.178 & 0.133 & 0.414 \\
$\beta_N$ & Estim. & 0.244 & 0.075 & 0.097 & 0.244 & 0.179 & 0.035 & 0.113 & 0.179 & 0.132 & 0.023 & 0.087 & 0.132 \\
      & P-value & 0.303 & 0.151 & 0.079* & 0.303 & 0.120 & 0.077 & 0.025* & 0.120 & 0.083* & 0.061 & 0.014* & 0.083* \\
\textbf{OTM-Put} &       &       &       &       &       &       &       &       &       &       &       &       &  \\
$\beta_P$ & Estim. & -3.941E-03 & 0.035 & -0.071 & -3.941E-03 & -0.059 & 0.023 & -0.103 & -0.059 & -0.038 & 0.018 & -0.072 & -0.038 \\
      & P-value & 0.788 & 0.148 & 0.450 & 0.788 & 0.409 & 0.182 & 0.127 & 0.409 & 0.486 & 0.204 & 0.159 & 0.486 \\
$\beta_N$ & Estim. & 0.216 & 0.037 & 0.144 & 0.216 & 0.191 & 0.024 & 0.144 & 0.191 & 0.226 & 0.020 & 0.188 & 0.226 \\
      & P-value & 0.058* & 0.045 & 8.781E-03** & 0.058* & 9.732E-03** & 0.011 & 7.331E-04*** & 9.732E-03** & 7.724E-05*** & 1.657E-04 & 1.076E-06*** & 7.724E-05*** \\
\end{tabular}%

}
\end{sidewaystable}

\end{document}